%% file: main.tex
\pdfoutput=1
\documentclass[manuscript]{acmart}
\usepackage{enumitem}
\usepackage{subfigure}
\usepackage{multirow}
\usepackage{array}
\usepackage{stfloats}
\usepackage{caption}
\usepackage{graphicx}
\usepackage[flushleft]{threeparttable}

\AtBeginDocument{%
  \providecommand\BibTeX{{%
    \normalfont B\kern-0.5em{\scshape i\kern-0.25em b}\kern-0.8em\TeX}}}

\acmConference[IPSN '23]{IPSN 2023: Conference on Information Processing in Sensor Networks}{May 9-12, 2023}{San Antonio, Texas USA}
\acmBooktitle{IPSN '23,
  May 9-12, 2023, San Antonio, Texas USA}
\newcommand{\proposal}{\textit{PointSplit}}
\newcommand{\rev}[1]{{\color{black}#1}}

\begin{document}

\title{PointSplit: Towards On-device 3D Object Detection with Heterogeneous Low-power Accelerators}


\author{Keondo Park}
\affiliation{%
  \institution{Graduate School of Data Science, Seoul National University}
  \city{Seoul}
  \country{Republic of Korea}
}
\author{You Rim Choi}

\affiliation{%
  \institution{Graduate School of Data Science, Seoul National University}
  \city{Seoul}
  \country{Republic of Korea}
}
\author{Inhoe Lee}
\affiliation{%
  \institution{Graduate School of Data Science, Seoul National University}
  \city{Seoul}
  \country{Republic of Korea}
}
\author{Hyung-Sin Kim}
\affiliation{%
  \institution{Graduate School of Data Science, Seoul National University}
  \city{Seoul}
  \country{Republic of Korea}
}

\input{sec/0_abstract.tex}



\keywords{3D object detection, On-device machine learning, Edge computing, Quantization}
\maketitle

\input{sec/1_intro.tex}
\input{sec/2_related_work.tex}

\input{sec/3_motivation.tex}
\input{sec/4_pointsplit.tex}
\input{sec/5_implementation.tex}
\input{sec/6_experiments.tex}
\input{sec/7_discussion.tex}
\input{sec/8_conclusion.tex}

\begin{acks}
This work was supported by Creative-Pioneering Researchers Program through Seoul National University. Hyung-Sin Kim is the corresponding author.
\end{acks}

\bibliographystyle{ACM-Reference-Format}
\bibliography{main}

\end{document}

%% file: sec/0_abstract.tex
\begin{abstract}
Running deep learning models on resource-constrained edge devices has drawn significant attention due to its fast response, privacy preservation, and robust operation regardless of Internet connectivity. While these devices already cope with various intelligent tasks, the latest edge devices that are equipped with multiple types of low-power accelerators (i.e., both mobile GPU and NPU) can bring another opportunity; a task that used to be too heavy for an edge device in the single-accelerator world might become viable in the upcoming heterogeneous-accelerator world.
To realize the potential in the context of 3D object detection, we identify several technical challenges and propose \proposal, a novel 3D object detection framework for multi-accelerator edge devices that addresses the problems. Specifically, our \proposal design includes (1) 2D semantics-aware biased point sampling, (2) parallelized 3D feature extraction, and (3) role-based group-wise quantization. 
%
%
%
We implement \proposal on TensorFlow Lite and evaluate it on a customized hardware platform comprising both mobile GPU and EdgeTPU. Experimental results on representative RGB-D datasets, SUN RGB-D and Scannet V2, demonstrate that \proposal~on a multi-accelerator device is \rev{24.7$\times$}  faster with similar accuracy compared to the full-precision, 2D-3D fusion-based 3D detector on a GPU-only device.

\end{abstract}

%% file: sec/1_intro.tex
\section{Introduction}

On-device machine learning (ML), which runs deep neural networks (DNNs) directly on an edge device (e.g., mobile phone), has drawn increased attention due to its potential to enable real-time and private ML applications. Development of low-power AI accelerators (e.g., mobile GPU and NPU), model compression schemes (e.g., quantization, pruning, and knowledge distillation), and system execution techqnies has enabled to run various intelligent tasks on a device, such as 2D object detection and language processing models~\cite{Howard_2019_ICCV, sun-etal-2020-mobilebert, Cai_Li_Yuan_Niu_Li_Tang_Ren_Wang_2021,choi2022scriptpainter,chen2018marvel,yi2020heimdall,apicharttrisorn2019frugal}.

\begin{figure}[t]
  \centering  
  \includegraphics[width=.7\linewidth, bb=0 0 800 300]{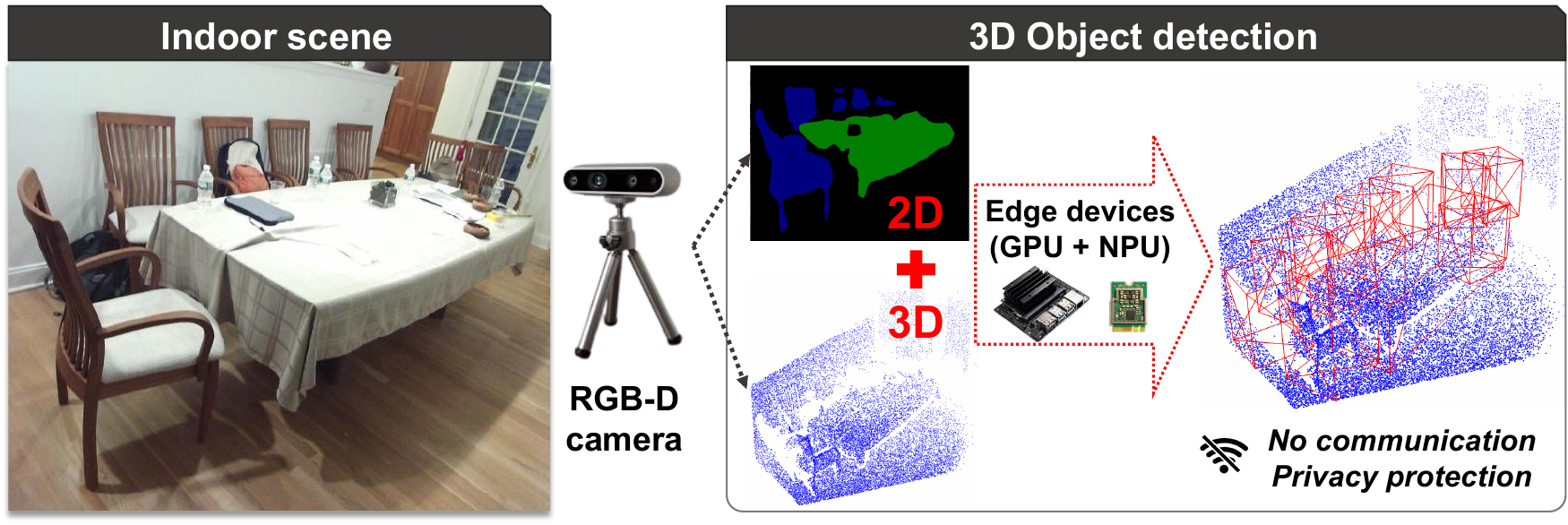}
  \vspace{-3ex}
   \caption{Target scenario: On-device 3D indoor scene understanding via RGB-D camera. On-device detection provides advantages on privacy, latency, and communication burden. 2D-3D fusion can improve detection accuracy while utilizing both GPU and NPU can accelerate on-device inference speed.}
   \Description{There is the target indoor scene on the left. RGB-D camera takes the snapshot of the scene to produce both 2D and 3D data. Edge devices which has GPU and NPU in it processes both 2D and 3D information to complete 3D object detection task. As a result chairs and a table are detected in the scene. At the right corner, the advantage of this system is described as "No communication" and "Privacy protection" with no-wifi icon.}
  \label{fig:application}
\end{figure}

Furthermore, although an edge device used to have a single type of AI processor, the recent emergence of heterogeneous processor System-on-Chips (SoCs)~\cite{8963950} has made the state-of-the-art mobile devices equipped with both high-end mobile GPU and NPU. The new class of edge devices with \textit{multi-type accelerators} present an opportunity to investigate interesting issues in the regime of on-device ML, such as intra-device parallelism and algorithm-system co-optimization by understanding different characteristics of the accelerators.  
With such evolution of low-power hardware, systems, and deep learning models together, more complex tasks that used to be far from resource-constrained devices, such as 3D object detection, might be able to run directly on device in real-time.
Specifically, running 3D object detection directly on resource-constrained devices, instead of powerful remote servers, has the potential to significantly expand the scope of AI applications. For example, as shown in Figure~\ref{fig:application}, a fast understanding of 3D indoor scenes directly on an edge device can be an important building block of the upcoming mixed reality~\cite{guan2022deepmix}. This work aims to investigate this new opportunity: \textit{on-device 3D object detection using both GPU and NPU}. 

\textbf{Challenges.} 
However, we identify that even with the latest edge devices containing both GPU and NPU, enabling on-device 3D object detection without sacrificing accuracy is  challenging in many ways: %
(1) 3D object detection is typically designed as a sequential process, making it hard to utilize GPU and NPU in parallel. 
(2) Since GPU and NPU have different strengths, a 3D object detection model should be analyzed thoroughly to distribute its computation to the two processors synergistically.
(3) Fusing 2D vision information with a 3D point cloud (e.g., using an RGB-D camera) can improve detection performance~\cite{Chen_2020_CVPR, qi2020imvotenet, xie2020mlcvnet} but makes the computational burden even heavier on the edge devices. 
(4) Quantization is necessary to reduce computation as well as to utilize NPU but given that 3D object detection is a sophisticated task, a naïve approach would significantly degrade the accuracy.

\vspace{1ex}\noindent
\textbf{Approach.} 
To tackle the challenges, we propose \proposal, a novel framework that provides system-driven model structure optimization for on-device 3D object detection. 
For the baseline deep neural network (DNN), we exploit VoteNet~\cite{qi2019deep}, a popular 3D object detection network based on the PointNet++ backbone~\cite{qi2017pointnet++}  for indoor 3D scene understanding, and borrow the idea of PointPainting~\cite{vora2020pointpainting} to augment features in a 3D point cloud (only geometric features) with 2D image semantics. 
Building on the baseline,  we devise three components for \proposal as below:
\begin{itemize}[leftmargin=*]
\item \textbf{2D semantics-aware biased 3D point sampling} aims to perform point sampling, a necessary process for processing a point cloud, more efficiently considering multi-type accelerator environments. To this end, we paint each 3D point using 2D image semantics and utilize the information to sample two complementary point sets, one from all points and the other more focused on the painted points (i.e., object-related points). We perform set abstraction (SA) process for the two point sets separately, called SA-normal and SA-bias, respectively. In this way, we generate two different views and perform two individual SA pipelines from a single 3D point cloud scene, which cooperate with each other to improve accuracy. 

\item \textbf{Parallelized 3D feature extraction} comes from the idea that widely used 3D point set abstraction methods~\cite{qi2017pointnet++}  comprise two operations, (1) point sampling and ball query that can be run only at GPU and (2) a neural net called PointNet~\cite{8099499} to process the sampled points that can be run at NPU. The two AI processors execute the two SA pipelines (SA-normal and SA-bias) interchangeably: GPU processes sampling and ball query for SA-normal while NPU processes PointNet for SA-bias, and vice versa. 

\item \textbf{Role-based group-wise quantization} is to compress neural networks without sacrificing accuracy and is motivated by the fact that layer-wise quantization significantly degrades accuracy while channel-wise quantization requires many quantization parameters. To find the sweet spot, we investigate each channel's weight and activation distribution and find out that the distribution heavily depends on the \textit{channel's role}. Based on the observation, we group channels according to their role and perform group-wise quantization, which preserves accuracy with only a few quantization parameters. 
\end{itemize}

\vspace{1ex}\noindent
\textbf{Contributions.} 
Our contributions can be summarized as follows:
\begin{itemize}[leftmargin=*]
  \item This work is the first to investigate on-device 3D object detection with heterogeneous low-power AI processors. Specifically, we propose \proposal, a novel framework that jointly designs system and algorithm to effectively reduce inference latency on resource-constrained devices.
  \item We deeply analyze the characteristics of a 2D-3D fusion-based 3D object detection model and design three unique components to reduce and parallelize computation without sacrificing accuracy: 2D semantics-aware point sampling, parallelized 3D feature extraction, and role-based group-wise quantization.  
  \item We implement VoteNet, a popular 3D object detection model, on TensorFlow from scratch\footnote{VoteNet and other state-of-the-art 3D object detection models are implemented on Pytorch (edge-unfriendly platform so far) but not on TensorFlow, which is a non-trivial entry barrier to research on-device 3D object detection. To the best of our knowledge, this work provides the first open implementation of VoteNet on TensorFlow.} and our \proposal~on TensorFlow Lite.\footnote{\rev{Code is available at \url{https://github.com/KeondoPark/votenet_tf} }} Furthermore, we build a test resource-constrained platform by combining NVIDIA Jetson Nano (including mobile GPU) and Google EdgeTPU (an NPU type).\footnote{\rev{This platform is a single device but not a system-on-chip (SoC). We expect performance improvement of \proposal when using a SoC including multi-type accelerators.}} 
  \item Experiments show that on two representative  datasets for indoor 3D object detection, SUN RGB-D~\cite{song2015sun} and Scannet V2~\cite{dai2017scannet},  \proposal~is up to \rev{24.7}  times faster than the full-precision, GPU-only baseline while providing similar accuracy.  
\end{itemize}


%% file: sec/2_related_work.tex
\vspace{-1ex}
\section{Related Work}

Given that this work is related to various fields, this section clarifies what techniques we leverage or are inspired by and what aspects our \proposal newly explores.

\vspace{-1ex}
\subsection{On-device Machine Learning} 

On-device ML refers to running deep neural network (DNN) inference locally without sending user data to the cloud. There has been a growing interest in on-device ML due to its advantages in latency and privacy. However, it is challenging to run DNN directly on edge devices because their memory, computational resource, and power consumption are strictly constrained. To address this problem, a number of lightweight DNN  architectures~\cite{Sandler_2018_CVPR, DBLP:conf/icml/TanL19, Tan_2019_CVPR, Tan_2020_CVPR} and model compression techniques~\cite{10.5555/2969239.2969366, DBLP:conf/iclr/FrankleC19, DBLP:journals/corr/HanMD15, Jacob_2018_CVPR, 44873} have been proposed. 
In addition, the development of low-power AI accelerators (e.g., mobile GPU and NPU) has enabled various DNN-based ML applications to be run on devices and showed notable results for some tasks~\cite{Howard_2019_ICCV, sun-etal-2020-mobilebert, Cai_Li_Yuan_Niu_Li_Tang_Ren_Wang_2021}. Furthermore, with the emergence of heterogeneous processor System-on-Chips (SoCs), scheduling or pipelining techniques have been developed to efficiently utilize multiple processors~\cite{9262933, 10.1145/3460352, 9525229}.

However, to our knowledge, there has not been any successful attempt for on-device 3D object detection even though heterogeneous low-power AI processors are given. As a step forward, this work presents a \textit{system-algorithm joint design} of 3D object detection to effectively reduce inference latency by fully leveraging the capacity of NPU and GPU on an edge device.

\vspace{-1ex}
\subsection{3D Object Detection} 

\rev{3D object detection is an essential component in robotics, AR/VR and autonomous driving, which require accurate 3D localization of objects. Here 3D localization includes measuring the distance between a user (or robot/vehicle) and an object and the size of the object (i.e., bounding box). For example, in AR/VR applications, inaccurate 3D localization cloud lead to unrealistic display of scenes or user dissatisfaction.}

Various methods have been proposed to estimate 3D bounding boxes of objects from point clouds. Many studies rely on voxel-based approaches to process 3D data, such as 3D CNN~\cite{7353481, 10.1145/3072959.3073608} and Voxel transformer~\cite{mao2021voxel}. 
To reduce the quantization error as well as large memory and computation cost inherent in voxel-based approaches, voxel feature encoding~\cite{Zhou_2018_CVPR}, hybrid voxel network~\cite{Ye_2020_CVPR}, or point-voxel fusion methods have been proposed~\cite{liu2019point, shi2020pv, shi2021pv}.
Another group of methods process point clouds directly for 3D scene understanding. PointNet \cite{8099499} and PointNet++ \cite{qi2017pointnet++} use symmetric functions to extract features from irregularly distributed points. VoteNet \cite{qi2019deep} exploits voting information from the features extracted from points by PointNet++. More recent work uses graph convolution to improve the feature extraction process \cite{chen2020hierarchical} or an enhanced voting scheme to improve detection accuracy \cite{zhang2020h3dnet}.

RGB information can be supplemented to understand 3D scenes. MV3D \cite{Chen_2017_CVPR} generates 3D object proposals from a bird's-eye view and uses deep fusion to combine 3D and 2D information. Frustum-PointNet \cite{qi2018frustum} utilizes 2D object detection results to guide 3D object detection.  3D-SIS~\cite{hou20193d} projects extracted features from 2D convolutions back to a 3D voxel grid to detect objects in a 3D scene. 
Given that these fusion techniques do not achieve expected performance improvement over 3D-only approaches, PointPainting \cite{vora2020pointpainting} proposes a sequential fusion as an alternative. It obtains 2D semantic segmentation scores and appends the information to each projected point in 3D space. Despite its advantage on accuracy, the sequential fusion significantly degrades latency.

In terms of 3D object detection model architecture, this work takes a point-based, 2D-3D fusion approach, inspired by VoteNet and PointPainting. With our design choices tailored for a multi-type accelerator environment, \proposal takes advantage of 2D-3D fusion to improve accuracy without sacrificing latency.

\vspace{-1ex}
\subsection{2D Semantic Segmentation} 

We utilize 2D semantic segmentation to fuse 2D image semantics with 3D point cloud to improve detection accuracy. In this regime, early work first suggested that convolutional neural network provides significant performance improvement over methods relying on hand-crafted features~\cite{long2015fully,drozdzal2016importance}. U-net~\cite{ronneberger2015u} proposed a U-shaped architecture to improve the capacity of the decoder by connecting expanding paths to contracting paths. 
Deeplab~\cite{chen2014semantic,chen2017deeplab,chen2017rethinking,chen2018encoder} further improved segmentation accuracy by using atrous convolution and a more advanced encoder-decoder structure. We use Deeplabv3+~\cite{chen2017deeplab} as our semantic segmentation network.

\vspace{-1ex}
\subsection{Deep Neural Network Quantization} 

Quantization is an active research area with the rising popularity of edge devices. It aims to carry out the inference with low-bit operations for the efficient use of resources while preserving accuracy. Ternary weight networks \cite{li2016ternary} or Binary Neural Networks~\cite{hubara2016binarized} binarize weights and activations of neural networks.  Jacob et al. \cite{Jacob_2018_CVPR} proposed an integer arithmetic only quantization scheme, which significantly accelerates inference and can run on accelerators that support only integer operations, such as EdgeTPU. In this work, we take the full quantization approach to run the model on EdgeTPU.

While most work on quantization targets image classification tasks,  a few recent studies~\cite{Chen_2021_CVPR, Li_2019_CVPR, ding2019req} suggest quantization techniques optimized for 2D object detection. To our knowledge, however, there has been no work that specifically targets the quantization of a \textit{3D object detector}. In doing so, we focus on quantization granularity, one of the key considerations in quantization. Layer-wise quantization~\cite{krishnamoorthi2018quantizing} determines the clipping range of the quantization from the statistics of the entire layer. On the other hand, statistics from each channel are used to calculate the clipping range in channel-wise quantization~\cite{huang2021codenet, Jacob_2018_CVPR}. Q-BERT \cite{shen2020q} groups multiple channels to decide the clipping range for quantizing the transformer network. Although our approach also groups multiple channels, we find out that doing it in a different manner is more effective for 3D object detection: taking model semantics into account, rather than grouping evenly.

%% file: sec/3_motivation.tex
\vspace{-1ex}
\section{Baseline and Motivation}

This section presents the baseline network for 2D-3D fusion-based 3D object detection that our \proposal builds upon, and analyzes the problems when naïvely applying the baseline for a multi-type accelerator environment, which motivates \proposal.

\vspace{-1ex}
\subsection{The Baseline: PointNet++ and PointPainting}

Our baseline is a 3D object detection model that fuses a 2D image and a 3D point cloud from an RGB-D scene. We choose VoteNet~\cite{qi2019deep} as the baseline 3D object detector, which is widely-used for indoor scene understanding. VoteNet utilizes PointNet++~\cite{qi2017pointnet++} as the backbone to extract features from a 3D point cloud. %
For 2D-3D fusion, we take the approach in PointPainting~\cite{vora2020pointpainting}, performing 2D semantic segmentation first and utilizing the semantic information for more accurate 3D object detection. We use Deeplabv3+~\cite{chen2017deeplab}  as the 2D semantic segmentation model and MobileNetV2~\cite{Sandler_2018_CVPR} as its lightweight feature extractor.

While the baseline sequentially runs Deeplabv3+ and VoteNet, its essence, highly related to our \proposal~design, is in the PointNet++ backbone and the fusion method in PointPainting, which are described below.

\vspace{1ex}\noindent
\textbf{PointNet++ for 3D Point Set Abstraction.}
Extracting meaningful features from a set of 3D points is important to detect objects from a 3D 
scene. While 2D image features  can be extracted purely with a neural net due to the dense nature of the RGB image, due to the sparse nature of 3D point clouds, it is essential for a 3D point set abstraction method to intermingle point manipulation with neural nets. 
To this end, PointNet++~\cite{qi2017pointnet++} has \textit{set abstraction (SA) layer} that includes both point manipulation and neural net.

Specifically, given a point cloud, an SA layer first constructs multiple groups of neighboring points by performing point sampling and ball query sequentially. To sample center point for each group, PointNet++ utilizes the farthest point sampling (FPS) method, which samples a new point that is most distant from  the already sampled points. Ball query draws a ball around each center point and groups neighboring points in each ball.
After the point manipulation, 
a local feature vector is extracted for each ball by processing a neural net called PointNet~\cite{8099499}. Since each ball is represented as its center point, the SA layer can be performed again based on the set of center points as a new point cloud input to extract higher-level features. PointNet++ repeats the SA layer four times to extract high-level features hierarchically from a raw-level point cloud.

\vspace{1ex}\noindent
\textbf{PointPainting for 2D-3D Fusion.}
Before processing a point cloud, PointPainting first performs semantic segmentation on a 2D image of the same scene, which divides the image pixels into two groups: foreground (object-related) and background groups. The semantic information is given to each 3D point as an additional feature. Then 3D object detection is performed based on the semantic-aware 3D point cloud, which improves accuracy.

\begin{figure*}[t]
  \centering
  \includegraphics[width=\linewidth, bb=0 0 1300 300]{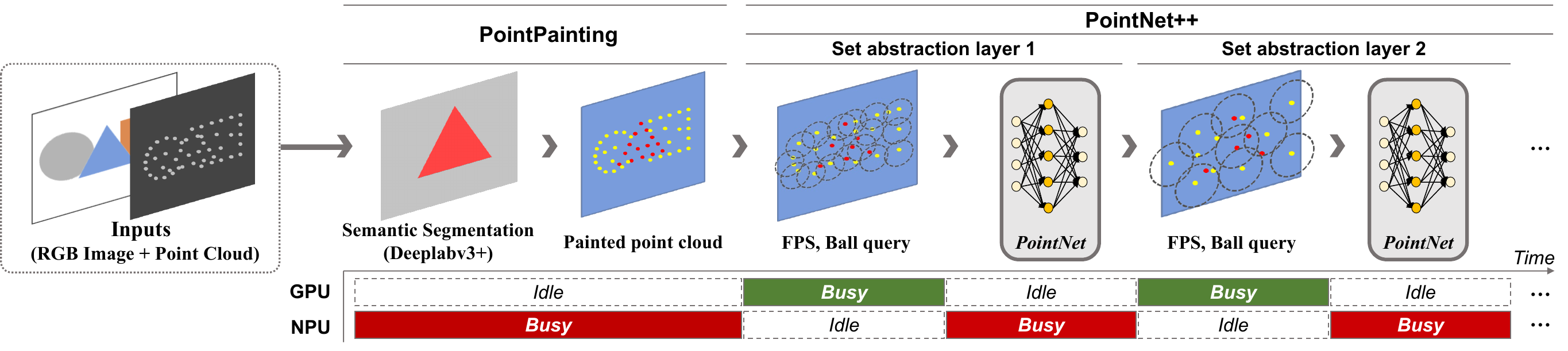}
   \vspace{-3ex} \caption{Illustration of naïve workload distribution to run the sequential pipeline of PointPainting on a GPU-NPU combined environment. Among the three figures in the input scene, only the triangle shape is assumed to be a valid object (foreground points). Either of the processors is always idle, waiting for the other to finish its job.} 
   \Description{At the very left, there are inputs(RGB Image + Point Cloud). At the right of inputs, the semantic segmentation is located.  Deeplabv3+ model is used for semantic segmentation. After semantic segmentation painted point cloud is obtained. Semantic segmentation and painted point cloud are grouped and termed as "PointPainting". After pointpainting, PointNet++ is executed. PointNet++ consists of multiple set abstraction layers. The set abstraction layer consists of FPS, Ball query and PointNet. Below the flow, corresponding busy processors(GPU and NPU) are highlighted. For pointpainting, GPU is idle and NPU is busy. In set absraction, GPU is busy and NPU is idle at FPS,Ball query. In contrast, GPU is idle and NPU is busy at PointNet.}
   \vspace{-2ex}
   \label{fig:pointnet2_naive}
\end{figure*}

\vspace{-1ex}
\subsection{Motivation: A Naïve Application of the Baseline on Multi-type Accelerators}

Running the fusion-based sequential 3D object detection pipeline on a GPU-only environment suffers from long  latency, which is also recognized in~\cite{vora2020pointpainting}. To mitigate the problem, the authors proposed a consecutive matching method, which reuses 2D segmentation results of a previous scene for detecting objects on the current scene. However, this approach is vulnerable to the difference between the current and previous scenes and cannot be applied to single-shot detection scenarios. By using GPU and NPU together, we aim to provide \textit{concurrent matching} that performs both 2D semantic segmentation and 3D object detection on the current scene.

When running the baseline on a multi-type accelerator environment (GPU and NPU), it is important to consider what operations can be executed on  NPU since it is faster than GPU but supports limited operations. 
As a neural network accelerator, NPU can process only Deeplabv3+ and PointNet, neither point sampling nor ball query. Therefore, to utilize both NPU and GPU, it is natural to perform point sampling and ball query on GPU, and PointNet and Deeplabv3+ on NPU. Figure~\ref{fig:pointnet2_naive} depicts this naïve approach.

As shown in Figure~\ref{fig:pointnet2_naive}, however, without changing the baseline's sequential process, the naïve workload distribution inevitably causes idle time on both processors. When processing PointNet++, NPU has to wait while GPU performs point sampling and ball query, and GPU also needs to wait while NPU processes PointNet. The same issue arises when fusing 2D and 3D information; while NPU performs 2D semantic segmentation via Deeplabv3+, GPU waits for the semantic segmentation results in the idle state. Although running these neural nets on NPU instead of GPU has its own speed gain, we aim to step further by reducing the idle time.  

%% file: sec/4_pointsplit.tex

\section{PointSplit}

This section presents our \proposal~design, which aims to answer the following questions: (1) Can we create two parallel SA pipelines to utilize both GPU and NPU simultaneously without sacrificing accuracy? 
(2) Can GPU do something meaningful using the point cloud while NPU processes 2D semantic segmentation? 
(3) How can we minimize accuracy drop when fully quantizing the baseline 3D object detector?

\begin{figure*}[t]
  \centering
    \includegraphics[width=\linewidth, bb=0 0 1300 400]{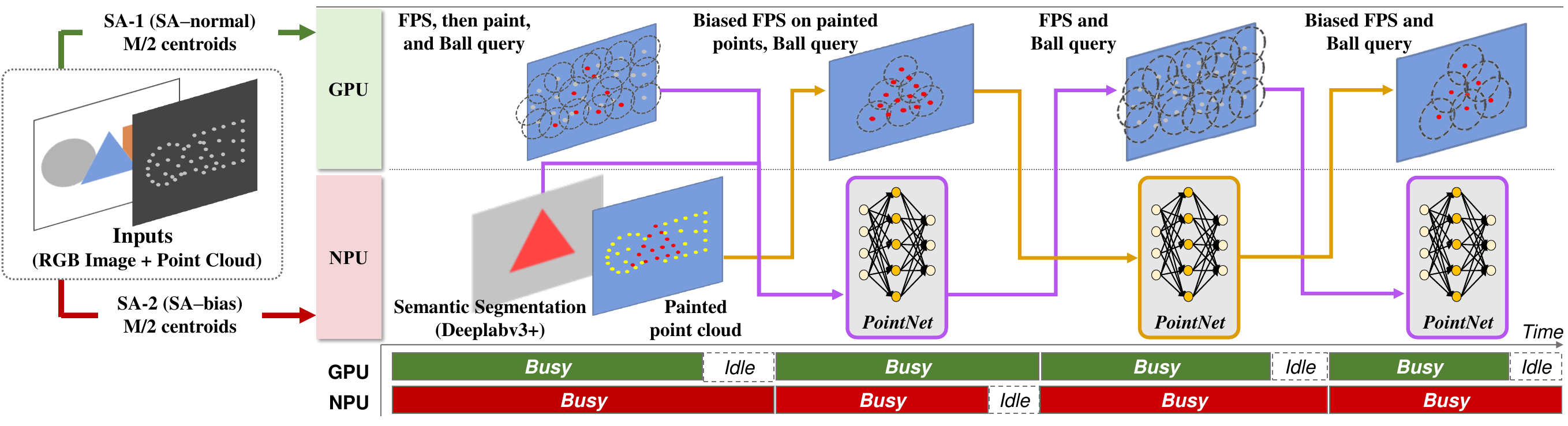}
    \vspace{-3ex}
  \caption{\rev{Illustration of \proposal's parallelized set abstraction (SA) pipeline. Each lightweight SA process in \proposal generates only half the number of balls compared to the conventional SA layers in PointNet++. When GPU processes point manipulation for an SA pipeline, NPU processes PointNet for the other SA pipeline in parallel, which reduces  idle time on each processor.}}
  \Description{At the very left, there are inputs(RGB Image + Point Cloud). At the right of inputs, two-row table is located. Top row corresponds to GPU and the bottom row corresponds to NPU. From the inputs, a green arrow goes to GPU, and named as "SA-1(SA-normal)" and M/2 centroids are used. Another arrow goes to NPU and named as "SA-2(SA-bias)" and M/2 centroids are used. The first item of GPU row is FPS, then paint and Ball query. Magenta arrow goes from the first item of GPU(FPS, then paint and Ball query) to PointNet which is at the second item of bottom row. The first item of the bottom row is semantic segmentation(Deeplabv3+) and Painted point cloud. This is connected to the second item of GPU via yellow arrow. The second item of GPU is "Biased FPS on painted points, Ball query". This is again connected to the third item of NPU via yellow arrow, which is PointNet. This is again connected to the fourth item of GPU which is Biased FPS and Ball query, via yellow line. The second item of NPU is connected to the third item of GPU which is FPS and Ball query via magenta line. This again connects to the fourth item of NPU which is PointNet. Below the table, corresponding busy processors are highlighted. Both GPU and NPU are almost always busy.}
  \label{fig:pipelining}
  \vspace{-2ex}
\end{figure*}

Figure~\ref{fig:pipelining} illustrates our parallel processing of the baseline network. To divide the SA process in PointNet++ into two \textit{lightweight parallel pipelines}, 
we design point sampling and ball query for each SA pipeline to generate  \textit{only half the number of balls} (i.e., the number of center points) while being processed on GPU. While NPU processes PointNet with the reduced number of balls for an SA pipeline (called SA-1), GPU performs point sampling and ball query again to generate the other half of balls for the other SA pipeline (called SA-2) in parallel. This method reduces computation for each SA pipeline and parallelizes point manipulation and neural net operations,  which reduces each processor's idle time. 

In addition, to utilize 2D semantic information for both lightweight SA pipelines without significant delay, one SA process (SA-1) \textit{jump-starts} on GPU without waiting for the segmentation results from NPU since the segmentation results are needed for PointNet, not point manipulation. After GPU and NPU finish the point manipulation (for SA-1) and 2D segmentation tasks, respectively, NPU computes PointNet for SA-1 by using the semantic information and GPU performs point manipulation for the other SA pipeline SA-2.

While the fundamental pipelining structure is effective in terms of latency, we aim to go further by applying different point sampling strategies for SA-1 and SA-2 to create synergy between the two for accuracy improvement. In addition, since the NPU-based acceleration is meaningful only when the object detection model can maintain accuracy after fully quantized, we develop a new quantization scheme for 3D object detectors.  


\vspace{-1ex}
\subsection{2D Semantics-aware Biased Point Sampling}
\label{sec:ps_biased}

To create synergy between the two lightweight SA processes (SA-1 and SA-2), we focus on the fact that SA-1 starts before the 2D segmentation task is finished but SA-2 starts after the 2D segmentation. This means that while both SA-1 and SA-2 utilize the semantic information when processing PointNet for feature augmentation, SA-2 can \textit{utilize the 2D semantic information also for its point manipulation}, if it is useful. Given that PointPainting utilizes the semantic information only for neural net operations, the idea of 2D semantics-aware point manipulation is new.

Specifically, since 2D semantic information distinguishes foreground points (those on valid 3D objects) from background points, we propose \textit{semantics-aware biased point sampling} by giving different priorities for foreground and background points when performing point sampling (FPS in the case of PointNet++). The intuition is that point sampling with a biased distribution can generate an augmented view \rev{for PointNet} from the same 3D scene, which improves the model's detection performance. 
\rev{Since a 3D input scene for PointNet consists of \textit{sampled points} instead of the whole point cloud, multiple different (augmented) inputs can be generated from an original point cloud scene depending on how the input points are sampled.}

To apply the semantics-aware biased sampling strategy for the FPS method, we manipulate the distance between two 3D points $p_1$ and $p_2$, denoted as $d(p_1,p_2)$, according to the type of the two points (foreground or background). Considering a point set $\mathcal{S}$ and its subset $\mathcal{A}$ ($\subset\mathcal{S}$) comprising the foreground points in $\mathcal{S}$, we re-define the distance metric $d(p_1,p_2)$ as follows:
\input{eq/biasedFPS}
Here $(x_1,y_1,z_1)$ and $(x_2,y_2,z_2)$ are the 3D coordinates of $p_1$ and $p_2$, respectively. In addition, $w_0$ is a weight coefficient  that can prioritize (when $w_0>1$) or de-prioritize (when $w_0<1$) foreground points in the FPS process.
For example when $w_0$ is larger than 1, 
the distance metric intentionally increases distance between $p_1$ and $p_2$ if at least one of them is included in \(\mathcal{A}\). If both points are in \(\mathcal{A}^c\), their distance is calculated normally. Thus, points in \(\mathcal{A}\) are more likely to be selected as the farthest point in each iteration of FPS.

\begin{figure}[t]
  \centering
  \includegraphics[width=.8\linewidth, bb=0 0 800 400]{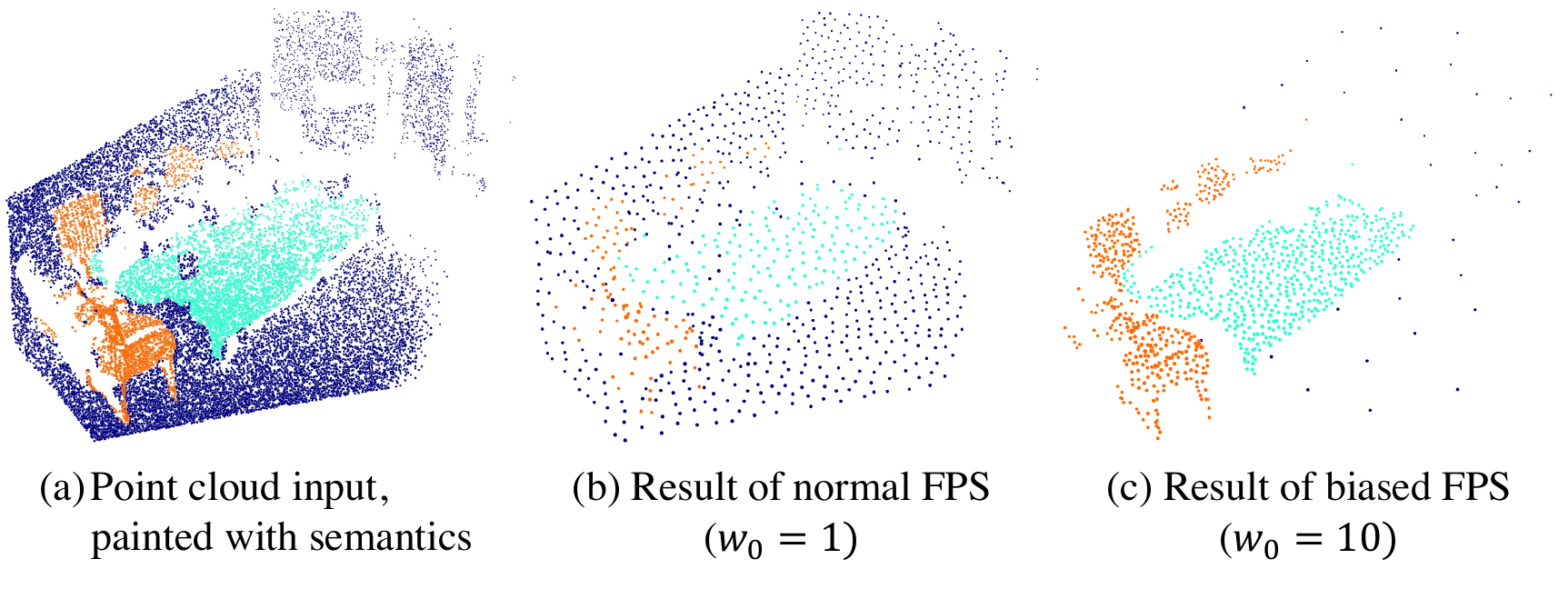}
  \vspace{-3ex}
  \caption{Illustration of \proposal's  semantics-aware biased point sampling. Using the same point cloud scene, our biased sampling can create significantly different multiple views by controlling the weight coefficient $w_0$.}
  \Description{There are three subfigures. The first one is "Point cloud input, painted with semantics". It is a point cloud of indoor point scene. There are four chairs and a table. The table is painted green and chairs are painted orange. Background is painted navy. The second subfigure is Result of normal FPS(w_0 equals to 1). This is sampled point cloud from the first subfigure. The points are sampled uniformly and the entire scene is mostly preserved. Similar to the first subfigure, the table is painted green, chairs are painted orange and the background is painted navy. The third subfigure is Result of biased FPS(w_0 equals to 10). More points are sampled from the table and chairs. Table and chairs are also painted with the same color as the first subfigure. Very few background points are sampled in this subfigure.}
  \vspace{-2ex}
  \label{fig:biasedFPS}
\end{figure}

Figure \ref{fig:biasedFPS} illustrates the impact of different $w_0$ values on the result of FPS. When $w_0=1$, points are sampled equally from both the foreground and background areas as the regular FPS does (Figure \ref{fig:biasedFPS}(b)). When a large weight is given to the painted (foreground) area ($w_0=10$), most points are sampled from the painted area (Figure \ref{fig:biasedFPS}(c)). The impact of $w_0$ value on the performance will be evaluated in Section~\ref{sec:expr_accuracy}. Overall, for a single point cloud input (Figure \ref{fig:biasedFPS}(a)), our biased sampling strategy can produce different multiple views.

\subsection{PointNet++ Optimization for Parallism}
\label{sec:pointnet_opt}

\begin{figure*}[t]
  \centering
    \includegraphics[width=\linewidth, bb=0 0 1400 300]{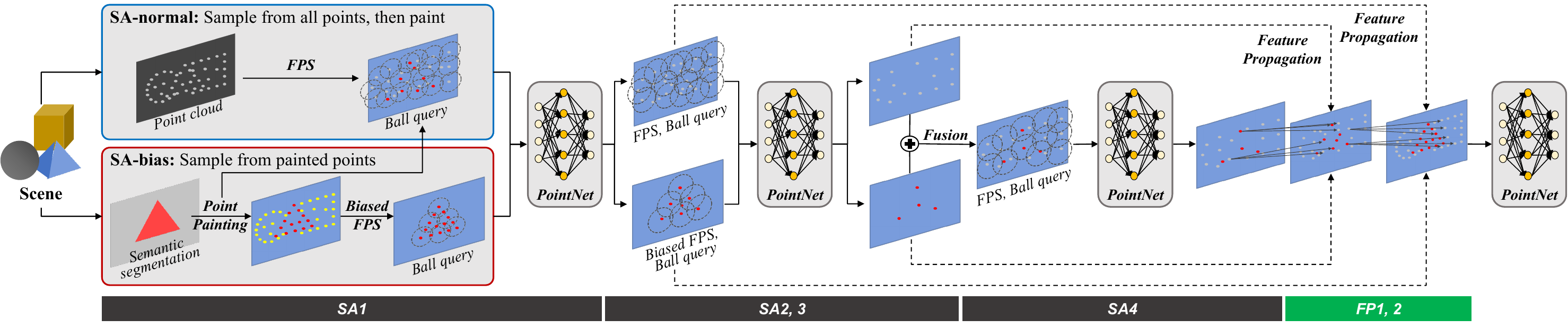}
    \vspace{-4ex}
  \caption{Illustration of PointNet++ structure optimized for \proposal. (1) An input point cloud is divided into two heterogeneous SA pipelines, one with regular FPS and the other with biased FPS. (2) The two SA pipelines share a single PointNet for data augmentation effect. (3) The two SA pipelines are merged before the fourth SA layer. (4) After SA layers, two FP layers are processed back to back and the last single PointNet produces the final output.}
  \Description{At the very left, there is a scene with multiple objects. One arrow goes to top then right to connect to the top box, and another arrow goes to bottom then right to connect to the bottom box. Top box is SA-normal: Sample from all points then paint. Inside the box, there is point cloud at left and right arrow named FPS connects to the sampled point cloud. Ball query is done over this point cloud. The bottom box is SA-bias: Sample from painted points. Inside the box, there is Semantic segmentation at left. The right arrow goes to painted point cloud. Another right arrow named Biased FPS goes to another point cloud. Ball query is done over this point cloud. Two arrows come out of the top box and bottom box and both goes to right to PointNet. From the beginning to this stage is marked as "SA1". Two arrows come out of PointNet. Top arrow goes to FPS, ball query and bottom arrow goes to Biased FPS and Ball query. Then, right arrow comes from both top and bottom and goes to PointNet. These operations are marked as "SA2"/"SA3". Again, two arrows come out of PointNet. Top arrow goes to FPS, ball query and bottom arrow goes to Biased FPS and Ball query. Two resulting point sets are fused together to produce the single point cloud. Right arrow comes from this point cloud and goes to pointnet. This is marked as SA4. From PointNet right arrow connects to the point cloud. Three arrows per each point is connected to the points in the right point cloud, which is more dense. This is termed as Feature propagation. The first feature propagation is connected from the point cloud at the end of SA3 via dashed arrow. Another feature propagation is repeated. This time, the dashed line is now connected from the point cloud at the end of SA1. Finally, the right arrow connects from the point cloud to PointNet at the very right.}
  
    \vspace{-2ex}
  \label{fig:2waySA}
\end{figure*}

With the two separate lightweight SA pipelines, SA-1 with regular sampling and SA-2 with biased sampling, we optimize the PointNet++ architecture to perform the two SA pipelines simultaneously on GPU and NPU, as illustrated in 
Figure \ref{fig:2waySA}. From now, we call SA-1 SA-normal and SA-2 SA-bias. 
Assume that an input point cloud for a regular SA layer has \(N\) points and $M(<N)$ centroids are sampled in the point manipulation stage.  
For SA-normal in \proposal, \(M/2\) centroids (\textit{half compared to the regular SA}) are sampled under regular FPS without using 2D semantic information. These centroids help the network to capture the overall context of the 3D scene. For SA-bias, another set of \(M/2\) centroids is sampled under biased FPS with more weight given to the foreground points. The biased point set contains more information for objects. We use $w_0=2$ for biased FPS on foreground points, which will be discussed in Section~\ref{sec:expr_accuracy}. 

We fine-tune the PointNet++ architecture to improve accuracy. Importantly, among the four SA layers\footnote{For the first three SA layers, the number of centroids for each normal SA and biased SA is 1024, 512, and 256, respectively. The radius for the ball query is 0.2, 0.4, 0.8, and 1.2 in each of the four SA layers, as in VoteNet.
} in PointNet++, the SA-bias pipeline uses biased FPS only for its first two SA layers; normal FPS is applied for the subsequent SA layers to capture the overall context at the end. %
In addition, the two sets of centroids from SA-normal and SA-bias are fused before the last (fourth) SA layer.
As for the neural network part (i.e., PointNet), we do not separately train two versions of PointNet for the two lightweight SA pipelines but train a single PointNet for both SA-normal and SA-bias. By doing so, we not only keep the network size from increasing but also expose the network to more diverse inputs with different characteristics, enabling robust detection (i.e., data augmentation effect). 

\input{tab/computaion_FP_layer}

Lastly, PointNet++ has two feature propagation (FP) layers after the four SA layers, each of which includes point manipulation and PointNet similar to an SA layer. For the two FP layers, we do not maintain parallel processing  because the two point sets are already fused before the fourth SA layer. In addition, we remove PointNet from each FP layer and attach a single shared fully-connected layer at the end of the second FP layer. As shown in Table~\ref{tab:computationFPLayer}, this simple modification achieves multiple advantages: reducing communication overhead between GPU and NPU, the number of model parameters 50.3\%, and the computation overhead 33.6\%. Despite the size and computation reduction from this change, we confirmed that it does not hurt the detection accuracy of the model.

%
%

\subsection{Role-based Group-wise Quantization}
\label{sec:role-quant}

Quantization is necessary to accelerate 3D object detection on an edge device but should be done carefully to not lose accuracy, given that 3D object detection is a complicated task. To this end, we carefully consider how to set quantization granularity. 
Various levels of granularity have been proposed, such as layer-, channel-, and group-wise quantization~\cite{krishnamoorthi2018quantizing, huang2021codenet, Jacob_2018_CVPR, shen2020q}. As can be inferred from the names, these techniques determine the clipping range for weights or activations depending on their distributions in an entire layer, each channel, or a group of several channels, respectively. 
Channel-wise quantization is the most sophisticated method, providing the best accuracy but requiring the largest number of quantization parameters. On the other hand, layer-wise quantization requires relatively small overhead but results in significant accuracy loss. Group-wise quantization is halfway between channel- and layer-wise quantization in terms of both the overhead and the accuracy loss. 
However, simply selecting one of the existing options might end up with inefficient quantization since model characteristics are not considered.


For accurate quantization using a small number of parameters, we observe  distributions of activations and weights in VoteNet, finding out that each channel's weight and activation distributions vary greatly in the last layer of voting and proposal modules.
We analyze the model structure and reveal that different distributions between groups of channels at a single layer come from their \textit{different roles}. Both the voting and proposal modules of VoteNet produce heterogeneous outputs that consist of  xyz-coordinates, features of the resulting points, bounding box size, etc. 
\rev{For example, the proposal module consists of different channels in charge of center regression, heading bin regression and classification, size regression and classification, and object classification. 
We observe that the distributions of weights and activations in channels appear similar according to their roles.}

\input{tab/channel_role.tex}

\begin{figure}[h]
  \centering
  \subfigure[Voting weights]{\includegraphics[width=.2\textwidth, bb=0 0 600 600]{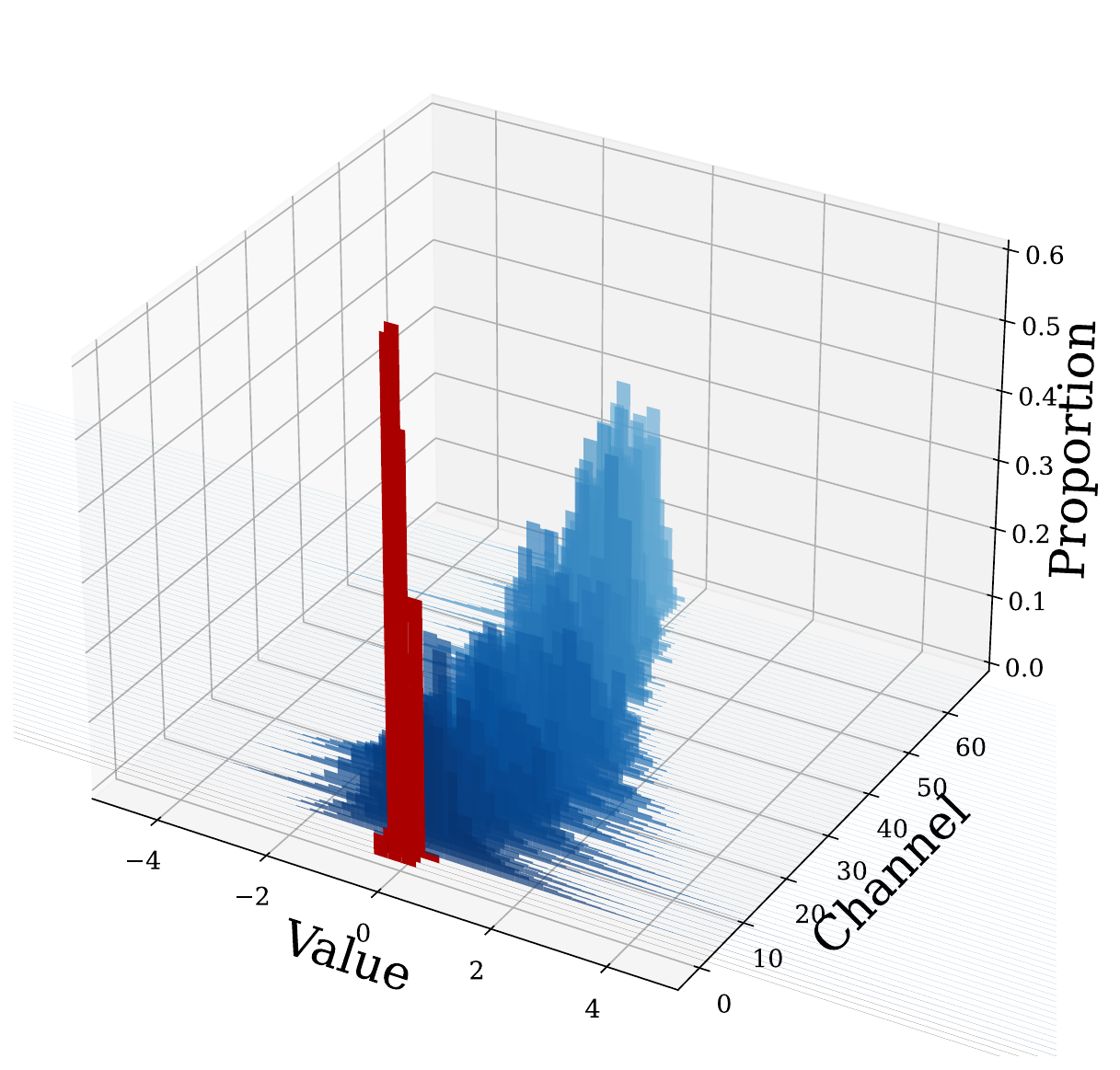}}
  \subfigure[Voting activations]{\includegraphics[width=.2\textwidth, bb=0 0 600 600]{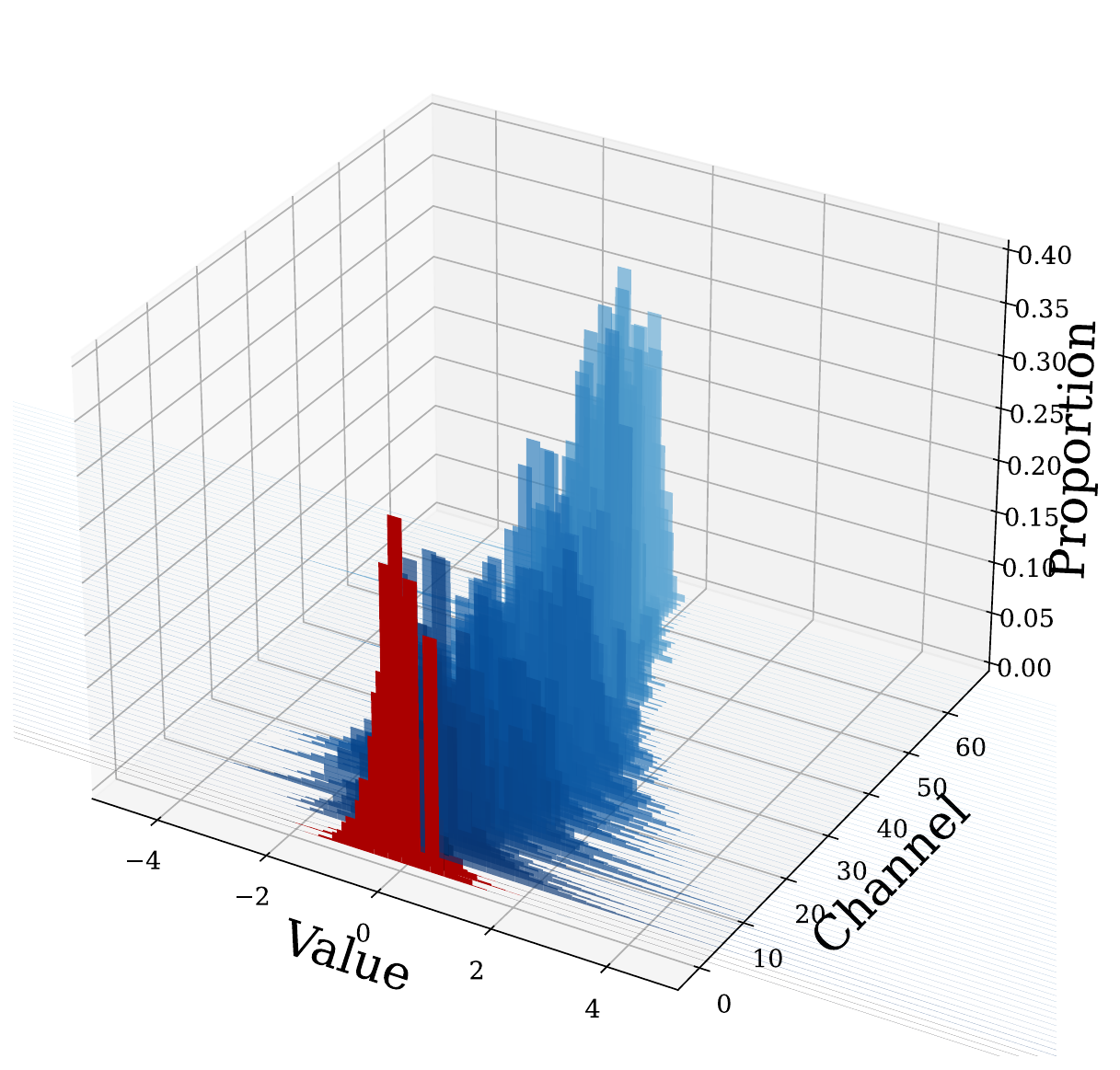}}
  \subfigure[Proposal weights]{\includegraphics[width=.2\textwidth, bb=0 0 600 600]{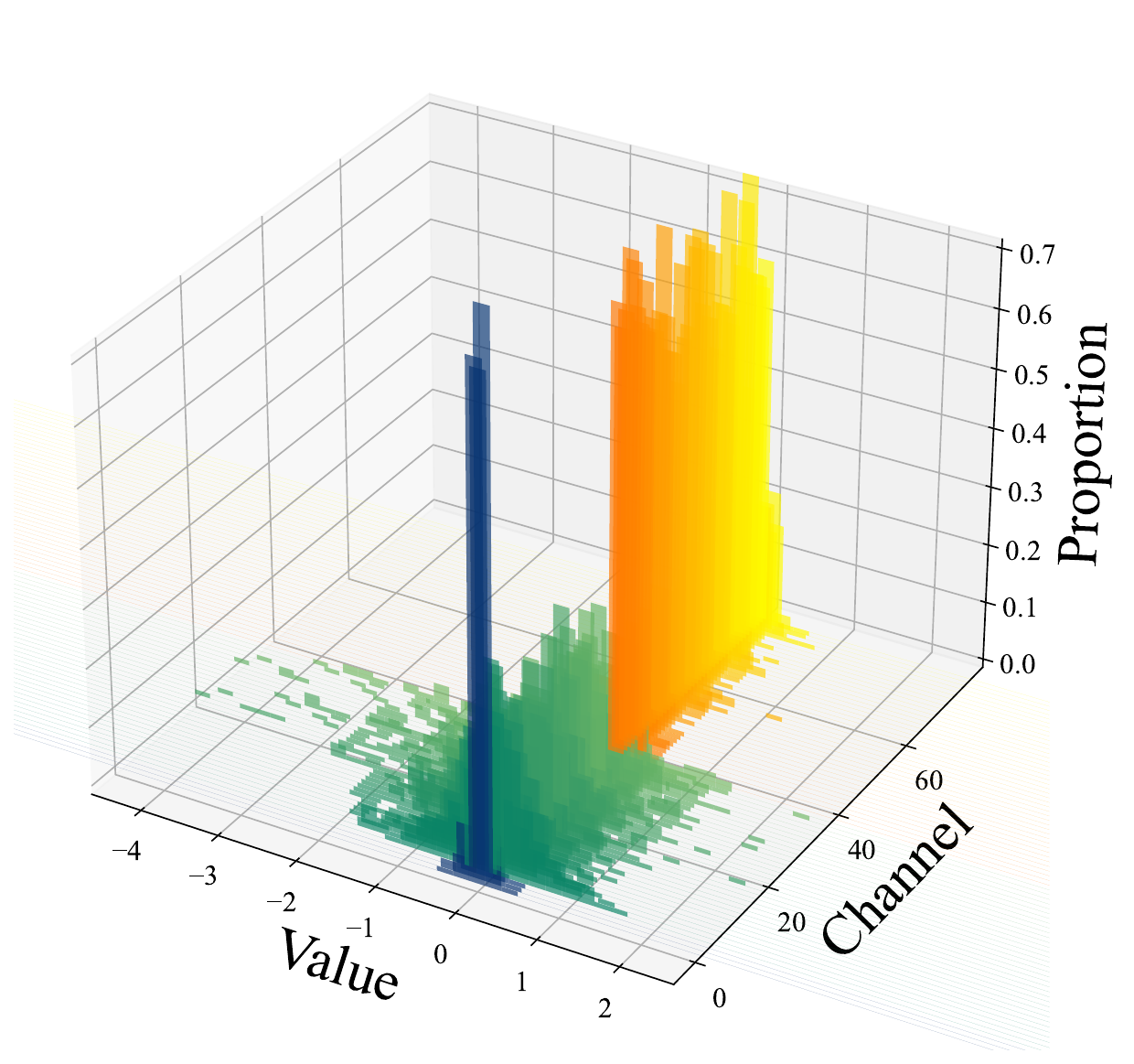}}
  \subfigure[Proposal activations]{\includegraphics[width=.2\textwidth, bb=0 0 600 600]{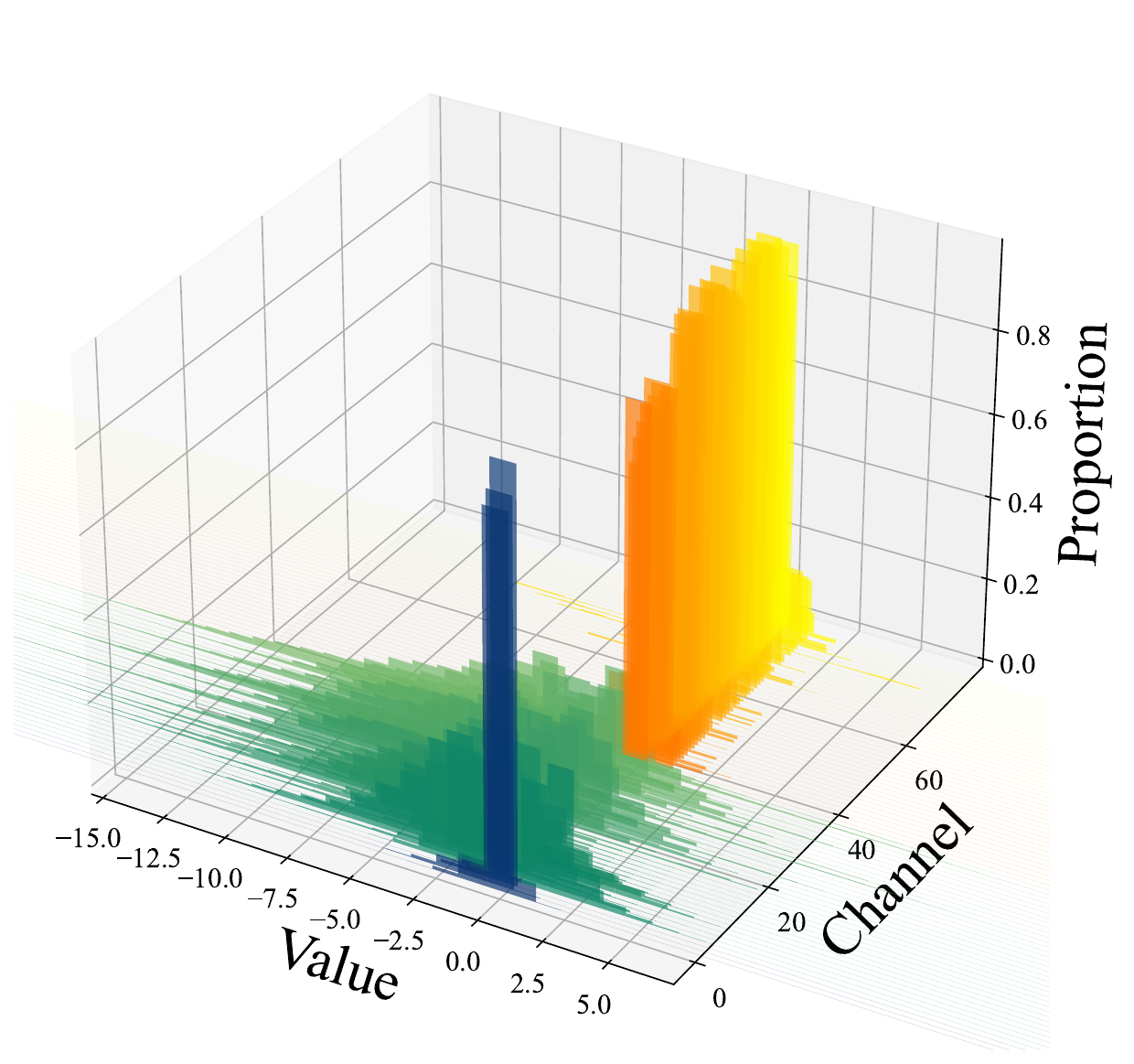}}
    \vspace{-3ex}
  \caption{\rev{The distributions of weights and activations in voting/proposal module. Channels in different group are marked as different color in the figure. Since there are too many channels for visualization in the voting module, the values in 4 consecutive channels are grouped as a single distribution.} }
  \Description{There are 4 3D charts. Across all charts, x-axis is value, y-axis is channel and z-axis is proportion. The first chart describes the distribution of Voting Weights. At each channel(y-axis), the distribution of value is represented as histogram in x-z plane. The distribution of the first three channel colored red is very densely distributed centered at 0. On the other hand, the distribution is more dispersed at other 64 channels and colored blue. The second chart is Voting activations. Similar shape is observed as the first chart in the second chart. THe third chart is Proposal Weights. the first three channel show centered distribution around 0 and colored blue. Next 34 channels are more dispersed and colored green. THe distribution of last 42 channels are more centered around 0 and colored yellow. THe fourth chart represents proposal activations and similar shape is observed as the third chart.}
  \label{fig:DistOfWnA}
    \vspace{-3ex}
\end{figure}

\rev{Importantly, we further discover that the distributions can be grouped according to whether the channel is responsible for classification or regression.} 
To utilize this characteristic, we group channels in the layer according to each channel’s role. 
In the voting module, channels are divided into two groups: the one in charge of predicting xyz-coordinates and the other for predicting the features. 
In the proposal module, channels are divided into three groups \rev{as shown in Table~\ref{tab:channel_role}}: the one in charge of predicting xyz-coordinates, another for heading bin, size cluster, and object classification, and the last group for regressing the size and orientation of the bounding box. 
We use post-training quantization~\cite{Jacob_2018_CVPR} to fully quantize the weights and activations to 8-bit integer.

\begin{figure}[H]
\centering
    \centering
    \includegraphics[width=.45\linewidth, bb=0 0 800 800]{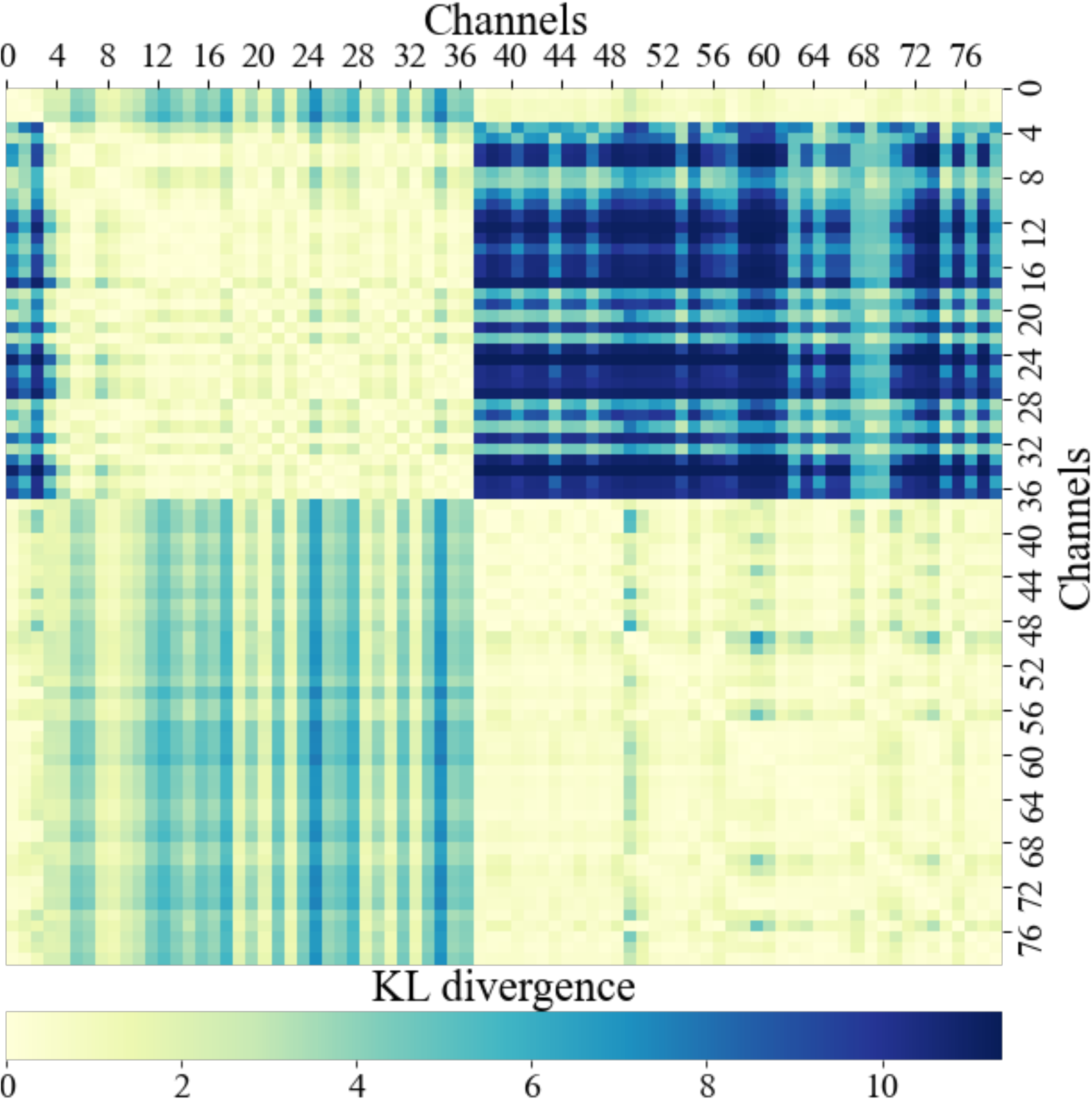}
    \vspace{-2ex}
    \caption{Kullback-Leibler (KL) divergence of activations in a proposal module of VoteNet. Dark blue implies larger KL divergence. KL divergence between different role-based channel groups has greater magnitude (e.g. group 3-27 vs 28-69).}\label{fig:KLdivergence}
    \Description{This is 79 by 79 heatmap. Light color represents low KL divergence and Dark color represents high KL divergence. Block diagonal of size 3 by 3, 34 by 34 and 42 by 42 are colored bright. But the color of other area is somewhat dark. The darkest area is the intersection of row 4-38 and column 38-79.}
    \vspace{-3ex}
\end{figure}

\rev{To clarify our role-based grouping, we  rearrange the order of the channels in both the voting and proposal modules in the last layer of VoteNet according to their role-based groups and plot the distributions of their weights and activations in Figure~\ref{fig:DistOfWnA}.}
\rev{The figure confirms} that each channel's weight and activation distributions vary greatly in the last layer of voting and proposal modules, depending on its role: which type of outputs to take charge of. 
For example, 
As shown in Figures~\ref{fig:DistOfWnA}(c) and \ref{fig:DistOfWnA}(d), for the first three channels of the proposal module (i.e., \rev{blue} bars, \rev{Group 1 in Table~\ref{tab:channel_role}}), weight and activation values are densely distributed around the mean value and min/max range is small. On the other hand, the next group of 24 channels (i.e., \rev{green} bars, \rev{Group 2 in Table~\ref{tab:channel_role}}) has a more dispersed distribution of weights and activations.
As another visualization, Figure \ref{fig:KLdivergence} shows Kullback-Leibler (KL) divergence of activations in a proposal module of VoteNet. The figure confirms that  distribution difference between activations from different channels is noticeable when channels are in different role-groups. 

%% file: eq/biasedFPS.tex
\begin{equation}
\label{eqn:biasedFPS}
    \begin{split}
    d(p_1, p_2) ~=~ &w * \sqrt{(x_1 - x_2) ^2 + (y_1 - y_2)^2 + (z_1 - z_2)^2 }, \\
    &\text{where } w=
        \begin{cases}
            w_0 & \text{if } p_1\in \mathcal{A}~ \text{or}~ p_2 \in \mathcal{A}   \\
            1 & \text{otherwise}
        \end{cases}
    \end{split}
\end{equation}

%% file: tab/computaion_FP_layer.tex
\begin{table}[t]
\centering
  \caption{Comparison of the amount of computation and model size between the feature propagation (FP) layers in PointNet++ and \proposal}
    \vspace{-3ex}
  \label{tab:computationFPLayer}
  \begin{tabular}{ccc}
    \toprule
    \multirow{2}{*}{Components} & \textbf{PointNet++} & \textbf{\proposal} \\
                                & Two PointNets & One modified PointNet\\
    \midrule
    \# of Parameters            & 398,336  & 197,888 \\
    MAdd                        & 304 M   & 202 M \\
    
  \bottomrule
\end{tabular}
\vspace{-3.5ex}
\end{table}

%% file: tab/channel_role.tex
\begin{table}[t]
\centering
  \caption{\rev{Three groups divided according to the role of channels in the proposal module of VoteNet}}
    \vspace{-2ex}
  \label{tab:channel_role}
\begin{tabular}{ccc}
\toprule
\textbf{Role-Group} & \textbf{Channels} & \textbf{\# of channels} \\
\midrule
Group1 & xyz-coordinates & 3 \\
\hline
\multirow{4}{*}{Group2} & Objectness score & 2 \\
                         & Heading bin classification & 12 \\
                         & Size classification & \# of classes \\
                         & Objectness category classification & \# of classes \\
\hline
\multirow{2}{*}{Group3} & Heading bin regression & 12 \\
                         & Size regression & \# of classes $\times$ 3 \\
\bottomrule
\end{tabular}
\vspace{-2ex}
\end{table}

%% file: sec/5_implementation.tex
\section{Implementation}
\label{sec:implementation}

\subsection{Datasets}

We train/test \proposal on two representative datasets for indoor 3D scene understanding: SUN RGB-D~\cite{song2015sun} and Scannet V2~\cite{dai2017scannet}.

\vspace{1ex}\noindent
\textbf{SUN RGB-D (the primary dataset).} 
Given that each SUN RGB-D image is a single RGB-D shot, SUN RGB-D is the primary dataset that fits our scenario in which an edge device performs inference on a single RGB-D shot. The SUN RGB-D dataset includes 10,335 RGB-D images taken indoors. 5,285 images are used for training and 5,050 images are used for validation. Segmentation annotations are provided for RGB images and 3D oriented bounding boxes of 37 categories are provided. We use the same data preparation step in VoteNet~\cite{qi2019deep} including conversion of the depth images into point clouds and data augmentation.

\vspace{1ex}\noindent
\textbf{Scannet V2 (the secondary dataset).} 
In contrast to SUN RGB-D, a Scannet V2 scene is not constructed from a single RGB-D shot but $\sim$1,500 shots with various different views that scan a $\sim$20$\times$ wider area more completely, resulting in much less occlusion and richer annotations. However, due to the scanning process, it takes a long time to get a scene for inference, which is \rev{not proper} for our scenario. Therefore, we utilize Scannet V2 as the secondary dataset.

Given that fusing 2D semantic information of all the 1,500 images is not practical, we randomly select \textit{only three 2D images} to evaluate the impact of 2D-3D fusion. The unbalanced information between 2D and 3D (i.e., using only three out of 1,500 2D images but a 3D point cloud containing all the 1,500 shots) is unfavorable for \proposal. The reason why we include Scannet V2 even though it is less practical and unfavorable for \proposal is to show that \proposal generally works well for multiple datasets.

Scannet V2 includes 1,513 scanned 3D indoor scenes and objects with  18 classes. 1,201 scans are used for training and 312 scans are used for validation. We also use RGB images and segmentation labels exported from the scanning stream for 2D-3D fusion.

\input{tab/votenet}

\vspace{-2ex}
\subsection{Platform with Multi-type Accelerators} 

\noindent\textbf{\rev{3D detector implementation on TensorFlow.}}
Although there are various DNNs that run on powerful servers, implementing them to run on an edge device is a labor-intensive and time-consuming task. Specifically, given that 3D object detection has been unexplored in the on-device ML regime, VoteNet is implemented only on Pytorch that is not an edge-friendly platform yet. To overcome the hurdle, we implement and train VoteNet on Tensorflow \textit{from scratch} and achieve comparable performance to the original Pytorch version~\cite{qi2019deep}, as shown in Table~\ref{tab:eval_votenetTF}. Thus, this work serves as the \textit{first open implementation} of VoteNet on TensorFlow, which can easily be converted into a TensorFlow Lite model to test on-device inference.
Specifically, we use Adam optimizer with an initial learning rate of 0.001. The learning rate is decreased by 10 times after 80 and 120 epochs. We train the model for 180 epochs and it takes around 10 hours on one RTX 3090 with Intel Xeon® Silver 4216 CPU on SUN RGB-D dataset~\cite{song2015sun} and 5 hours on Scannet V2 dataset~\cite{dai2017scannet}.

\begin{figure}[t]
\centering
        \includegraphics[width=.45\linewidth, bb=0 0 600 500]{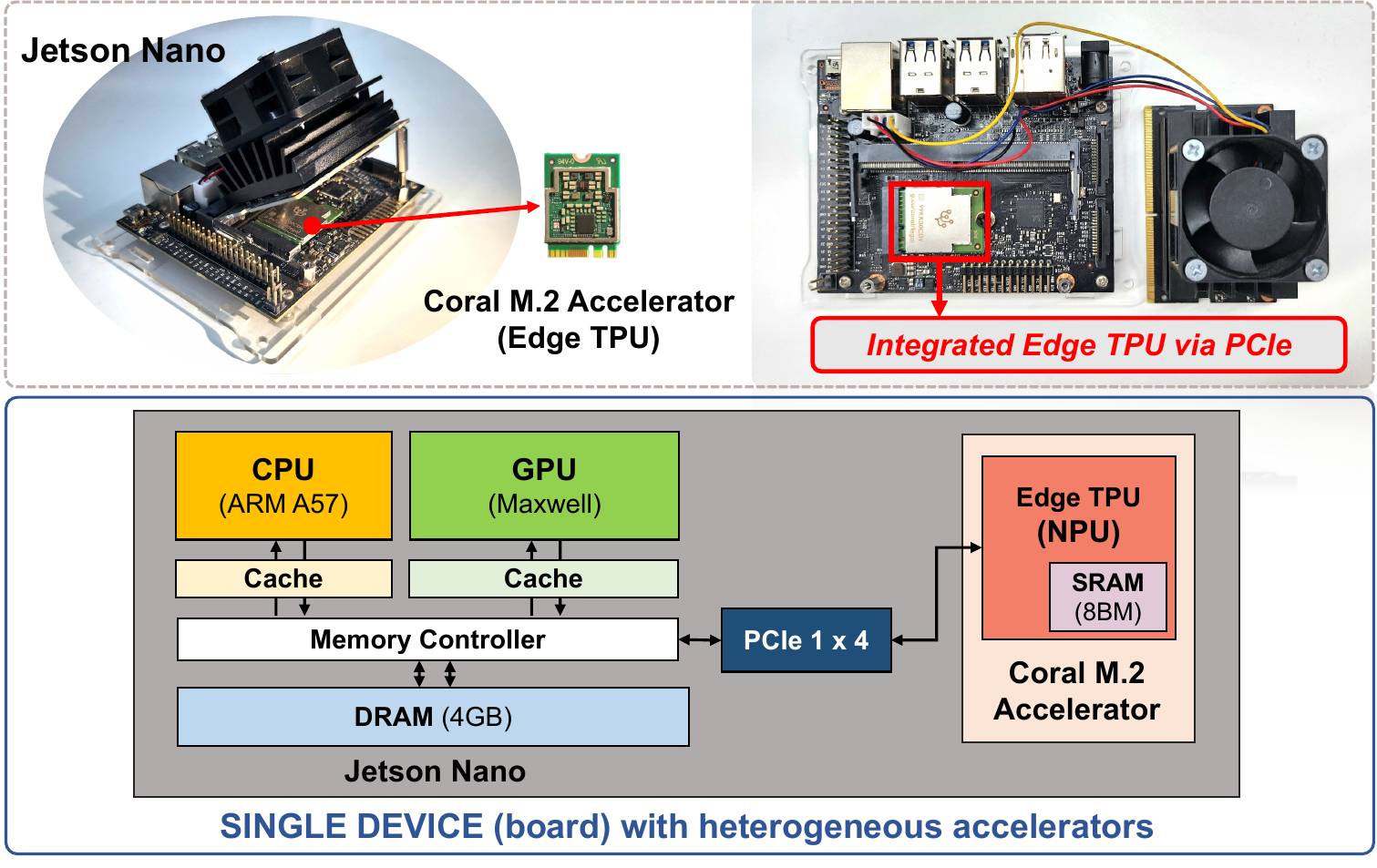}
\vspace{-3ex}
        \caption{\rev{Test environment with heterogeneous processors. We use NVIDIA Jetson Nano (GPU) equipped with Google Coral M.2 accelerator (EdgeTPU).}}\label{fig:LuxMeaDevices}
        \Description{At the top left, there is a photo of Jetson Nano equipped with Coral M.2 Accelerator(EdgeTPU). At the top right, there is another photo of Jetso Nano at different angle. Coral M.2 Accelerator is now more clearly visible. The red box is located around the Coral M.2 Accelerator and described as "Integrated Edge TPu via PCIe". At bottom the overall system architecture of the implementation  system is described. The outmost gray box represents Jetson Nano. At the left side in the gray box, CPU, GPU, CPU cache, GPU cache, memory controller and DRAM(4GB) is located. DRAM is at most bottom. Above DRAM memory controller is connected to DRAM. Memory controller is connected to CPU cache and GPU cache repsectively. CPU cache connects to CPU (ARM A57) and GPU cache is connected to GPU (Maxwell). At the center of Jetson Nano PCIe 1 x 4 connects memory controller and Coral M.2 Accelerator at right. Inside Coral M.2 Accelerator EdgeTPU (NPU) is located. At right bottom of Edge TPU, SRAM(8MB) is located.}
    \vspace{-3ex}
\end{figure}

\vspace{1ex}
\noindent\textbf{\rev{Hardware platform}.} 
To measure inference speed, we build a low-power platform with multi-type accelerators by combining Google Coral M.2 accelerator with NVIDIA Jetson Nano, as shown in Figure~\ref{fig:LuxMeaDevices}. NVIDIA Jetson Nano includes a quad-core ARM A57 CPU, 128-core Maxwell GPU of 512 GFLOPS and 4 GB 64-bit LPDDR4 memory. Coral M.2 accelerator includes an EdgeTPU coprocessor - an ASIC chip built for neural network inference, which is capable of 4 trillion operations per second. Coral M.2 accelerator is connected to Jetson Nano via PCIe Gen2 x 1 \rev{and shares its main DRAM memory}. 
Given that Coral EdgeTPU only supports integer operations, we quantize our model to INT8 and convert our model into TensorFlow Lite to compile it to be EdgeTPU-compatible. 
\rev{Note that although this platform is a single device including both GPU and EdgeTPU, it would be more ideal to utilize an integrated SoC as technology evolves. For example, although not available when we started this work, Apple's recent M1 architecture is designed for CPU, GPU, and NPU to share cache or memory.}

\vspace{-1ex}
\subsection{Deeplabv3+ for 2D Semantic Segmentation} 

To implement PointPainting, we use Deeplabv3+~\cite{chen2017deeplab} with MobileNet V2 backbone~\cite{Sandler_2018_CVPR} as a semantic segmentation network. We first pre-train Deeplabv3+ on COCO dataset~\cite{lin2014microsoft} and fine-tune the weights on each target dataset, SUN RGB-D and Scannet V2.
For fine-tuning for a target dataset, we use images and semantic segmentation labels in the target dataset along with COCO dataset. We  oversample some under-represented classes 5 times for the model to better locate those classes, as proposed in \cite{Kisantal_2019}. The oversampled classes include desk, dresser, night stand, bookshelf, bathtub from SUN RGB-D and window, bookshelf, picture, counter, desk, curtain, shower curtain, garbage bin from Scannet V2. We use SGD optimizer with momentum 0.9 and initial learning rate 0.05, and decay learning rate 0.94 on every \rev{2,000} training steps.
The final mIoU on SUN RGB-D validation images is 40.7\%, and the final mIoU on ScannetV2 validation images is 47.7\%.
Detailed per-class accuracy on both datasets are summarized in Tables \ref{tab:deeplab_sunrgbd} and \ref{tab:deeplab_scannet}, respectively.

\input{tab/deeplab_results_sunrgbd}
\input{tab/deeplab_results_scannet}



%% file: tab/votenet.tex
\begin{table*}[t]
\centering
  \caption{Per-class accuracy (mAP) at IoU threshold 0.25 of two VoteNet implementations on SUN RGB-D: (1) the original Pytorch implementation~\cite{qi2019deep} and (2) our own TensorFlow implementation. Our implementation provides comparable performance to the original version, serving as an open implementation that can be converted into TensorFlow Lite for on-device inference.}
  \vspace{-3ex}
  \label{tab:eval_votenetTF}
\resizebox{\linewidth}{!}{ 

  \begin{tabular}{ccccccccccccc}
    \toprule
    Item & Bathtub & Bed & Bookshelf & Chair & Desk & Dresser & Nightstand & Sofa & Table & Toilet & \vline & Overall\\
    \midrule
    VoteNet-Pytorch~\cite{qi2019deep} & \textbf{74.4} & 83.0 & \textbf{28.8} & \textbf{75.3} & 22.0 & 29.8 & \textbf{62.2} & \textbf{64.0} & 47.3 & \textbf{90.1} & \vline & \textbf{57.7} \\

    VoteNet-TensorFlow (ours) & 72.4 & \textbf{84.0} & 25.3 & 74.1 & \textbf{24.2} & \textbf{30.0} & 61.4 & 61.6 & \textbf{49.7} & 86.8 & \vline & 56.9 \\
  \bottomrule

    \end{tabular}
} 
\vspace{-3ex}
\end{table*}

%% file: tab/deeplab_results_sunrgbd.tex
\begin{table*}[t]
\centering
  \caption{Semantic segmentation accuracy (mIoU) of Deeplabv3+ on 2D images in the SUN RGB-D validation dataset.}
    \vspace{-3ex}
  \label{tab:deeplab_sunrgbd}

  \begin{tabular}{ccccccccccccc}
    \toprule
    Item & Bathtub & Bed & Bookshelf & Chair & Desk & Dresser & Nightstand & Sofa & Table & Toilet & \vline & Overall\\
    \midrule
    mIoU                 & 34.4 & 50.4 & 17.9 & 55.9 & 16.9 & 25.8 & 22.2 & 41.4 & 40.1 & 54.9 & \vline & 40.7 \\
  \bottomrule
    \end{tabular}
\vspace{-2ex}
\end{table*}

%% file: tab/deeplab_results_scannet.tex
\begin{table*}[t]
\centering
  \caption{Semantic segmentation accuracy (mIoU) of Deeplabv3+ on 2D images in the Scannet V2 validation dataset.}
  \vspace{-3ex}
  \label{tab:deeplab_scannet}
\resizebox{\linewidth}{!}{ 
  \begin{tabular}{ccccccccccccccccccccc}
    \toprule
    Item & Cab & Bed & Chair & Sofa & Table & Door & Wind & Bkshf & Pic & Cntr & Desk & Curt & Fridg & Showr & Toil & Sink & Bath & Gbg & \vline  & Overall\\
    \midrule
    mIoU                 & 45.8 & 53.2 & 50.8 & 55.4 & 60.6 & 40.7 & 25.5 & 20.7 & 28.5& 28.1 & 39.5 & 43.2 & 54.3 & 45.4 & 79.3 & 54.4 & 66.5 & 35.1 & \vline & 47.8  \\
  \bottomrule

    \end{tabular}
} 
\vspace{-2ex}
\end{table*}

%% file: sec/6_experiments.tex
\section{Experiments}

\subsection{Experimental setup}
\label{sec:expr_setup}

Following the recent practice in VoteNet, we use mean average precision (mAP) at 0.25 IoU threshold as our evaluation metric. The evaluation result is reported on the 10 most common categories for SUN RGB-D validation data, and 18 object categories for Scannet V2 validation data. As in VoteNet, we do not consider the bounding box orientation for Scannet V2 evaluation. As mentioned in Section~\ref{sec:implementation}, to fuse 2D information with a 3D point cloud scene, we use a single RGB image for the SUN RGB-D dataset and three RGB images for the Scannet V2 dataset. In addition, according to the standard practice for each dataset, we randomly sample 20,000 points and 40,000 points from an original point cloud of SUN RGB-D and Scannet V2, respectively, to construct an input point cloud for the first SA layer of PointNet++. Given that a Scannet V2 scene covers nearly 20 times larger area than a SUN RGB-D scene, an input point cloud in Scannet V2 is sparser than that in SUN RGB-D. 
Note that the results in SUN RGB-D is more important since it fits our scenario, while those in Scannet V2 show \proposal's general applicability.

We measure the latency on Jetson Nano equipped with EdgeTPU. We do three warm-up runs and experiment 20 times, then report averaged latency. Per each experiment, the latency is measured to process four 3D scenes and averaged to report per-scene latency.

\input{tab/eval_sunrgbd_class_25}
\subsection{Detection Accuracy}
\label{sec:expr_accuracy}

\noindent
\textbf{Analysis on SUN RGB-D (primary dataset).}
Table~\ref{tab:evalSunrgbdClass_25} shows per-class detection accuracy (mAP) of various 3D object detectors on the SUN RGB-D dataset.
VoteNet relies only on a point cloud without fusing 2D information, providing the lowest accuracy. PointPainting, the baseline network, combines Deeplabv3+ and VoteNet to fuse 2D semantic information for 3D object detection, which significantly improves accuracy over VoteNet (+3.3 mAP). This clearly shows the advantage of 2D-3D fusion for 3D object detection.

Interestingly, although our \proposal (full precision) originally focuses on efficient pipelining for multi-type accelerator environments, it ends up with \textit{even better accuracy} compared to PointPainting (+1.2 mAP). Specifically, out of 10 classes in SUN RGB-D, \proposal (full precision) achieves the best accuracy for 4 classes and the second best accuracy for other 4 classes. This verifies the effectiveness of our semantics-aware biased point sampling. By building two separate lightweight SA pipelines that have different views of a single 3D scene, with regular FPS and biased FPS, respectively, and making both pipelines pass through the same PointNet, \proposal trains PointNet more robustly  with data augmentation.

For comparison, we also test an ablated version of \proposal, called RandomSplit, which randomly divides an entire point set into two sets and applies regular FPS for both SA pipelines. Without biased sampling, RandomSplit does not provide an augmented view, resulting in similar accuracy to PointPainting.
Lastly, when \proposal is quantized with 8-bit precision, accuracy is  dropped marginally due to our role-based group-wise quantization. As a result, \proposal, even after quantized, performs comparably to the baseline PointPainting (full precision).

\input{tab/pipeline}

\vspace{1ex}\noindent
\textbf{Analysis on multiple datasets.} 
To show \proposal's accuracy gain more generally, Table \ref{tab:pipeline} summarizes accuracy performance of various schemes on both SUN RGB-D and Scannet V2 datasets, before and after quantization.
After quantized (layer-wise), both VoteNet and PointPainting experience remarkable performance degradation, which verifies our claim: activation and weight distributions in a single layer are too different to quantize using a single parameter set. In contrast, our \proposal (INT8) improves performance with very large margins (up to +30.6 mAP\rev{@0.25}) compared to both VoteNet and PointPainting in both datasets and performs even comparably to \proposal (FP32). This demonstrates the effectiveness of our role-based group-wise quantization scheme.

Although detailed trends are different due to different scene characteristics, the results in Scannet V2 also confirm that both semantics-aware biased point sampling and role-based group-wise quantization scheme significantly contribute to performance improvement. 
Specifically, we observe that RandomSplit degrades accuracy compared to PointPainting (-1.2 mAP\rev{@0.25}) but \proposal~recovers performance successfully. Given that point representation in Scannet V2 is already sparse (much sparser than that in SUN RGB-D, as mentioned in Section~\ref{sec:expr_setup}), sampling only half the number of points for each SA pipeline should be done carefully to not lose accuracy. The performance gap between RandomSplit and \proposal verifies the validity of our biased point sampling in this aspect.

\input{tab/groupfree3d.tex}
\vspace{1ex}\noindent
\textbf{Analysis on recent, heavy 3D object detectors.} 
\rev{The latest state-of-the-art 3D object detectors on the SUN RGB-D and Scannet V2 leaderboards adopt heavy and edge-unfriendly transformer architectures~\cite{yang2022boosting, wang2022multimodal,ran2022surface,liu2021group}. 
However, it is valuable to evaluate the effectiveness of our biased point sampling and parallel pipelining when applied to these models in terms of accuracy. 
To this end, we implement recent transformer-based GroupFree3D~\cite{liu2021group} and RepSurf~\cite{ran2022surface} on TensorFlow and apply PointPainting, RandomSplit, and \proposal to the two heavy baselines.\footnote{\rev{GroupFree3D employs a PointNet++ backbone and a transformer-based detection head~\cite{liu2021group} while RepSurf improves the input representation of GroupFree3D~\cite{ran2022surface}. Both models have been re-implemented and trained on TensorFlow, leveraging the hyperparameters of their respective Pytorch implementations. As a result, the TensorFlow-based models achieved a lower mAP compared to their original counterparts. It is worth noting that identifying optimal hyperparameters for TensorFlow could potentially improve their performance, which is out of the scope of this paper.}} Given that this evaluation is not for efficiency, we do not apply quantization and utilize two PointNets at the FP layers again (i.e., excluding the optimization in Table~\ref{tab:computationFPLayer}) when implementing \proposal to focus on better accuracy. 
The results in Table~\ref{tab:groupfree3d} demonstrate that \proposal successfully improves mAP when combined with GroupFree3D and RepSurf on both SUN RGB-D and Scannet V2 using two parallel SA pipelines. This finding confirms that \proposal is a viable technique for improving the accuracy of multiple 3D object detectors.}


\vspace{1ex}\noindent
\textbf{Deeper look into biased point sampling.} 
%
We analyze the impact of detailed design choices for the semantics-aware biased point sampling on performance. 
To this end, Table \ref{tab:bfps_weights} shows \proposal performance on SUN RGB-D with varying $w_0$ value. As mentioned in Section~\ref{sec:ps_biased}, as $w_0$ increases, the biased point sampling mechanism selects more foreground points than background points. The results in Table~\ref{tab:bfps_weights} show that as $w_0$ increases, mAP performance first increases but decreases again. Specifically, the highest accuracy is achieved when $w_0=2$. The results show that proper balance between foreground and background points is important when constructing an augmented scene via the biased point sampling. Specifically, sampling more foreground points turns out to be beneficial but sampling too many foreground points is detrimental.

\input{tab/bfps_sa}

\begin{table}[t]
    \centering
    \caption{Accuracy of \proposal on SUN RGB-D when the semantics-aware biased point sampling is applied to various SA layers.}
    \vspace{-2.5ex}
    \label{tab:bfps_sa}        
    \begin{tabular}{cc}
        \toprule    
        SA layers with biased FPS & mAP \\
        \midrule
        SA1 only & 60.4 \\
        SA1 and SA2 & \textbf{61.4} \\
        SA1, SA2 and SA3 & 60.1 \\
        All SA layers & 60.8 \\    
        \bottomrule        
    \end{tabular}
    \vspace{-3ex}
\end{table}

Next, we evaluate another design choice for the biased point sampling: which SA layers to apply the biased point sampling among the four SA layers in PointNet++. 
To this end, Table \ref{tab:bfps_sa} shows \proposal performance on SUN RGB-D when our biased sampling technique is applied to various SA layers, from the first (SA1) to the last (SA4). The results show that applying the biased point sampling to the first two layers provides the best performance but applying it to more SA layers causes performance degradation again.
Given that applying the biased point sampling to multiple layers consecutively increases the bias level, the results verify again the need for balancing foreground and background points to maximize \proposal performance.

\vspace{1ex}
\noindent\textbf{Impact of quantization methods.} 
\input{tab/quant_error}
Table \ref{tab:quantError} evaluates \proposal on the two datasets with varying  quantization granularity: layer-wise, group-wise, channel-wise, and our role-based group-wise methods. For the group-wise method, we group the entire layer into 2 (for the voting module) or 3 (for the proposal module) groups of an equal number of channels without considering their roles.

The results show that both layer-wise and group-wise quantization methods suffer from significant quantization errors but the channel-wise method incurs only marginal errors. This verifies that channels in a single layer of a 3D object detector have very different weight and activation distributions, which requires fine-grained quantization. The channel-wise quantization, however, is inefficient since it requires more than \rev{1,300} parameters to quantize a single layer.   
On the other hand, our role-based group-wise quantization achieves the sweet spot. It requires the same number of quantization parameters compared to the naïve group-wise quantization, 67$\times$ and 71$\times$ less than that of the channel-wise quantization on SUN RGB-D and Scannet V2, respectively. %
With such small number of parameters, our scheme dramatically improves accuracy over the group-wise quantization (+33.6 mAP on SUN RGB-D), and provides similar mAP compared to the heavy channel-wise quantization. 
This demonstrates the tight relationship between each channel's value distribution and its role in a 3D object detector.

\begin{figure}[t]

  \centering
  \vspace{-1ex}
  \subfigure[Latency on SUN RGB-D]{
    \includegraphics[width=.95\textwidth, bb=0 0 1300 600]{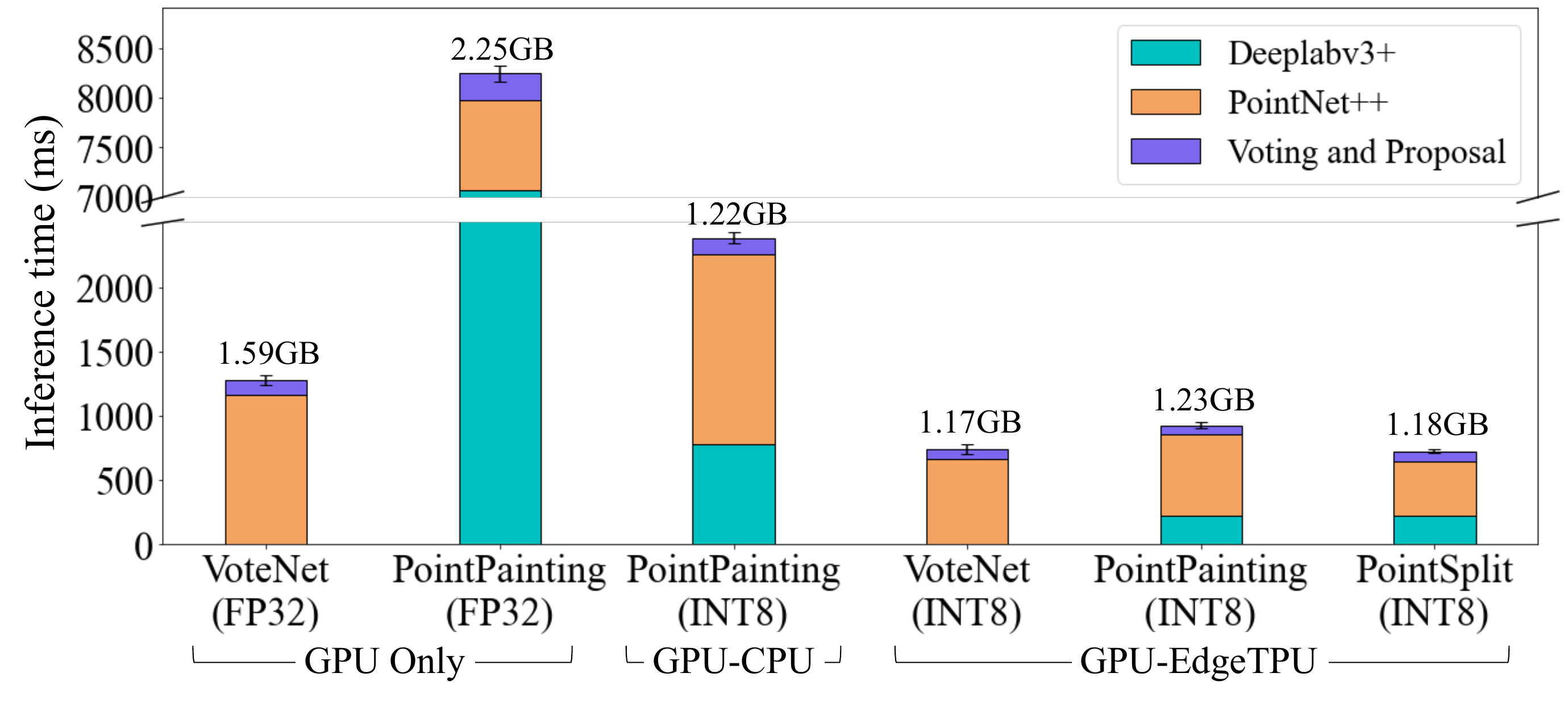}
    } 
  \subfigure[Latency on Scannet V2]{  
    \includegraphics[width=.95 \textwidth, bb=0 0 1300 600]{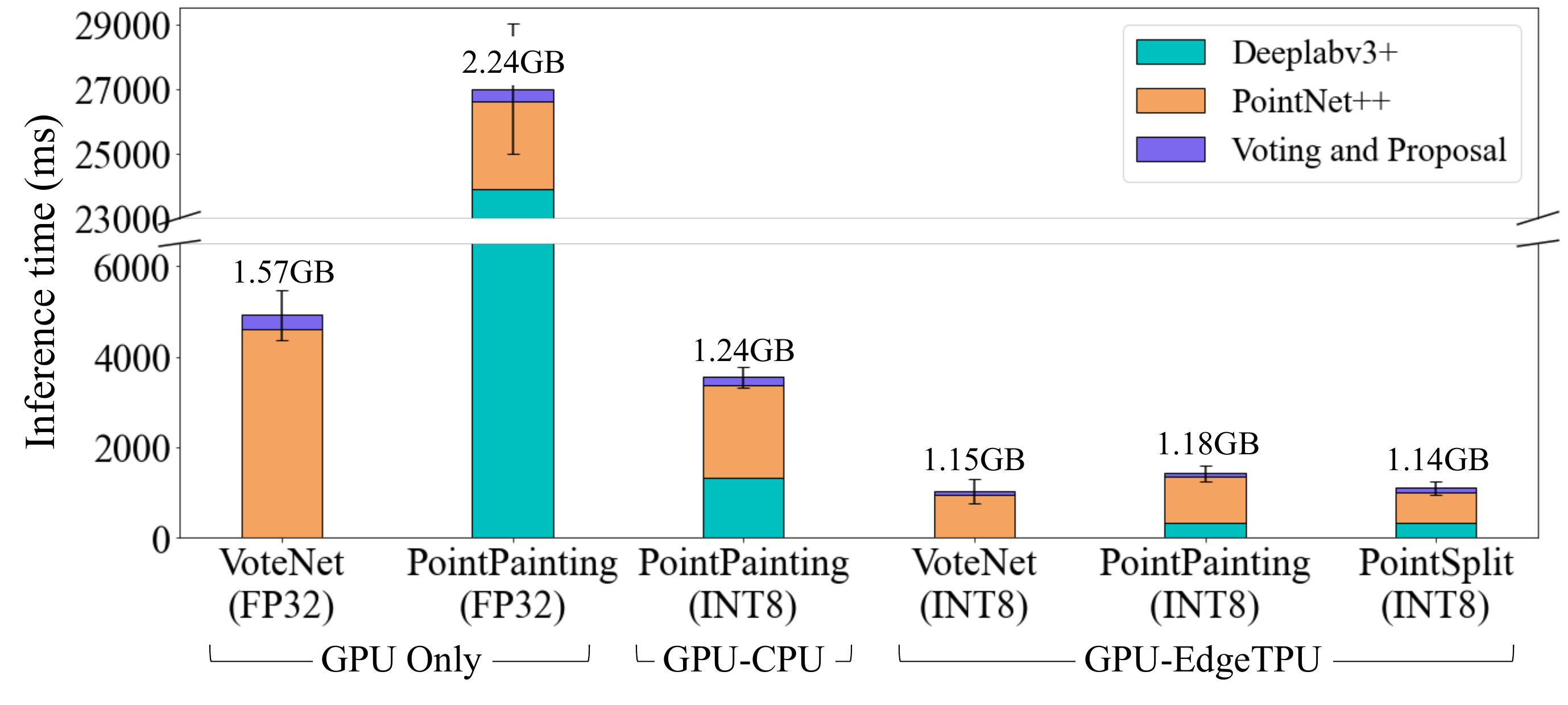}
    }
  \vspace{-3ex}
  \caption{\rev{Per-scene latency and peak memory of 3D object detectors. Latency on Scannet V2 is longer than that on SUN RGB-D due to more 3D points. Compared to running PointPainting (FP32) only on GPU, \proposal (INT8) is faster 11.4 times on SUN RGB-D and 24.7 times on Scannet V2.}}
  \Description{There are two subfigures. The first subfigure shows the latency on SUN RGB-D. There are 6 stacked bar charts - VoteNet(FP32), PointPainting(FP32), PointPainting(INT8), VoteNet(INT8), PointPainting(INT8), PointSplit(INT8). VoteNet(FP32) and PointPainting(FP32) are grouped together as "GPU Only". PointPainting (INT8) is grouped as "GPU-CPU". Remaining three items are grouped as "GPU-EdgeTPU". 
  The first item (VoteNet(FP32)) is two  bars stacked. The bottom one is PointNet++ and the top one is Voting and Proposal. The bottom bar is 1164ms and top bar is 112ms. 
  The second item(PointPainting(FP32)) is three bars stacked. The bottom one is DeeplabV3+, the middle one is PointNet++ and the top one is Voting and Proposal. The bottom bar is 7072ms, middle bar is 901ms and the top bar is 268ms. 
  The third item(PointPainting(INT8)) is three bars stacked. The bottom one is DeeplabV3+, the middle one is PointNet++ and the top one is Voting and Proposal. The bottom bar is 775ms, middle bar is 1483ms and the top bar is 126ms.
  The fourth item (VoteNet(INT8)) is two  bars stacked. The bottom one is PointNet++ and the top one is Voting and Proposal. The bottom bar is 667ms and top bar is 74ms. 
  The fifth item(PointPainting(INT8)) is three bars stacked. The bottom one is DeeplabV3+, the middle one is PointNet++ and the top one is Voting and Proposal. The bottom bar is 222ms, middle bar is 931ms and the top bar is 73ms.
  The sixth item(PointSplit(INT8)) is three bars stacked. The bottom one is DeeplabV3+, the middle one is PointNet++ and the top one is Voting and Proposal. The bottom bar is 227ms, middle bar is 414ms and the top bar is 83ms. 
  At the top of each item, the peak memory is written. 1.59GB, 2.25GB, 1.22GB, 1.17GB, 1.23GB and 1.18GB of peak memory is observed per each item.
  The second subfigure shows the latency on Scannet V2. There are 6 stacked bar charts - VoteNet(FP32), PointPainting(FP32), PointPainting(INT8), VoteNet(INT8), PointPainting(INT8), PointSplit(INT8). VoteNet(FP32) and PointPainting(FP32) are grouped together as "GPU Only". PointPainting (INT8) is grouped as "GPU-CPU". Remaining three items are grouped as "GPU-EdgeTPU". 
  The first item (VoteNet(FP32)) is two  bars stacked. The bottom one is PointNet++ and the top one is Voting and Proposal. The bottom bar is 4601ms and top bar is 324ms. 
  The second item(PointPainting(FP32)) is three bars stacked. The bottom one is DeeplabV3+, the middle one is PointNet++ and the top one is Voting and Proposal. The bottom bar is 23908ms, middle bar is 2708ms and the top bar is 389ms. 
  The third item(PointPainting(INT8)) is three bars stacked. The bottom one is DeeplabV3+, the middle one is PointNet++ and the top one is Voting and Proposal. The bottom bar is 1318ms, middle bar is 2048ms and the top bar is 180ms.
  The fourth item (VoteNet(INT8)) is two  bars stacked. The bottom one is PointNet++ and the top one is Voting and Proposal. The bottom bar is 945ms and top bar is 84ms. 
  The fifth item(PointPainting(INT8)) is three bars stacked. The bottom one is DeeplabV3+, the middle one is PointNet++ and the top one is Voting and Proposal. The bottom bar is 332ms, middle bar is 1056ms and the top bar is 80ms.
  The sixth item(PointSplit(INT8)) is three bars stacked. The bottom one is DeeplabV3+, the middle one is PointNet++ and the top one is Voting and Proposal. The bottom bar is 327ms, middle bar is 673ms and the top bar is 95ms. 
  At the top of each item, the peak memory is written. 1.57GB, 2.24GB, 1.24GB, 1.15GB, 1.18GB and 1.14GB of peak memory is observed per each item.  }
  \vspace{-3ex}
  \label{fig:Latency}
\end{figure}

\subsection{System Performance}

\noindent\textbf{Latency analysis.} 
Figure~\ref{fig:Latency} shows average latency for single-scene inference of various schemes, measured on our platform comprising Jetson Nano and Coral.
Generally, latency on Scannet V2 is longer than that on SUN RGB-D. This is because an input point cloud is twice larger and Deeplabv3+ runs three times more for a Scannet V2 scene than a SUN RGB-D scene, as in Section~\ref{sec:expr_setup}.
When using GPU only, PointPainting significantly increases latency  compared to VoteNet, requiring more than 8 and 27 seconds in SUN RGB-D and Scannet V2, respectively; despite its accuracy gain, 2D-3D fusion is not a viable option on classic resource-constrained devices.

In our platform including both GPU and EdgeTPU, however, the landscape can be shifted. First of all, running the point manipulation operation on GPU and the PointNet part on EdgeTPU significantly reduces latency, which shows the effectiveness of EdgeTPU that is optimized for neural net operations. In addition, although 2D-3D fusion using PointPainting increases latency on the multi-accelerator platform, the efficient pipelining scheme in  \proposal nullifies the slowdown in SUN RGB-D, resulting in inference speed comparable to VoteNet (INT8) with significantly better accuracy (+30.6 mAP@0.25, as in Table~\ref{tab:pipeline}). 
Compared to running PointPainting (FP32) only on GPU, running \proposal (INT8) on both GPU and EdgeTPU provides \rev{11.4$\times$ and 24.7$\times$} faster inference in SUN RGB-D and Scannet V2, respectively. 
Overall, the results suggest that in the upcoming multi-type accelerator era, 2D-3D fusion-based 3D object detection, which used to be a complex task, can run on an edge device without notable latency degradation.   

\vspace{1ex}
\noindent\textbf{Peak memory analysis.} 
\rev{Figure~\ref{fig:Latency} also shows peak memory usage of each scheme. Note that while VoteNet (FP32) and PointPainting (FP32) run on TensorFlow, other four schemes run on TensorFlow Lite. Since TensorFlow Lite does not support CUDA, the GPU-only environment utilizes TensorFlow.

PointPainting (FP32) on TensorFlow consumes more than 2.2 GB memory, which is one of the reasons why its latency is significantly high. We evaluate another version of PointPainting (INT8) on TensorFlow Lite by running the point manipulation operation on GPU and the PointNet (INT8) and Deeplabv3+ (INT8) part on CPU (i.e., GPU-CPU combination). The results show the impact of using a lightweight software platform: running neural nets on CPU with TensorFlow Lite is much faster and requires much less memory than on GPU with TensorFlow. 
Lastly, VoteNet (INT8), PointPainting (INT8), and \proposal (INT8) that run on the GPU-EdgeTPU environment and TensorFlow Lite consume similar peak memory. This verifies that compared to VoteNet and PointPainting, \proposal's parallel operation of GPU and EdgeTPU does not sacrifice memory for boosting inference speed.}

\begin{figure}[t]
  \centering
  \vspace{-2ex}
  \subfigure[Latency on SUN RGB-D]{\includegraphics[width=.48\textwidth, bb=0 0 1300 600]{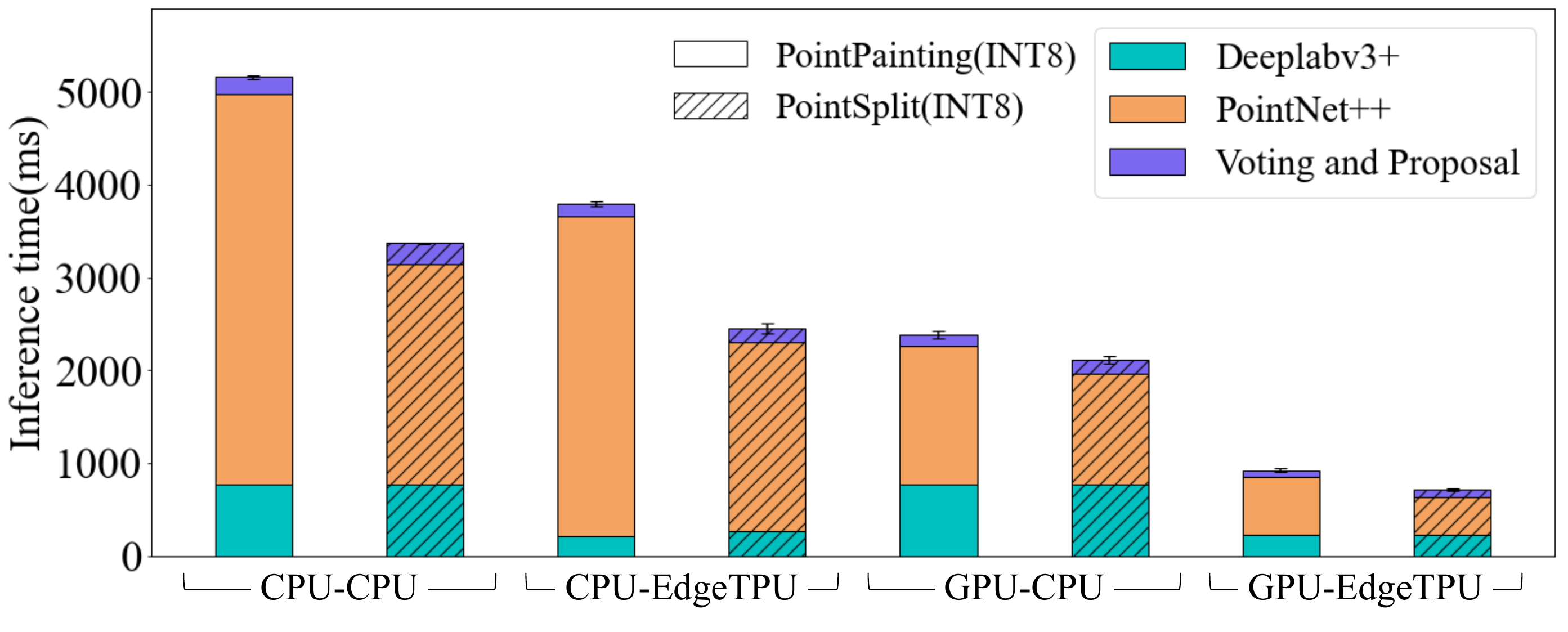}}
  \subfigure[Latency on Scannet V2]{\includegraphics[width=.48\textwidth, bb=0 0 1300 600]{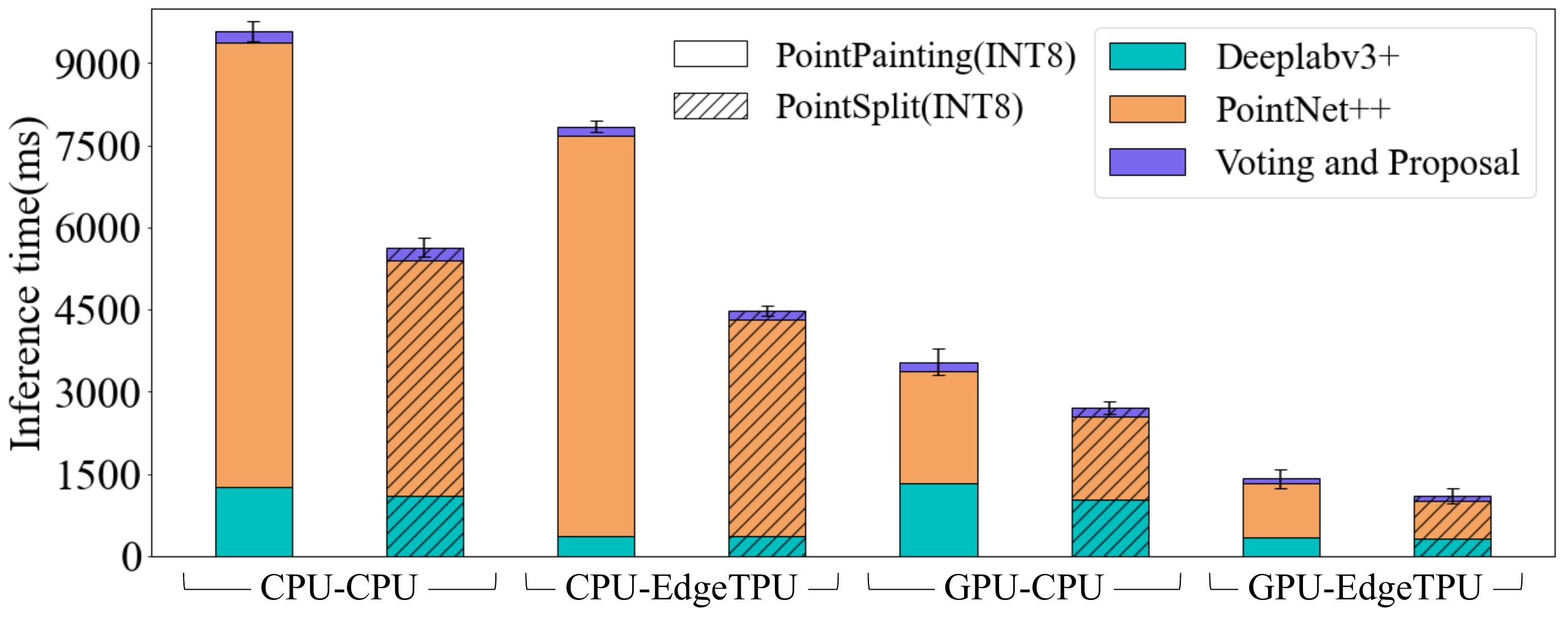}}
  \vspace{-3ex}
  \caption{\rev{Per-scene inference latency of VoteNet-based PointPainting (INT8) and \proposal (INT8) on various combinations of multiple processors. \proposal reduces latency regardless of hardware configurations.}}
  \Description{ There are two subfigures. The first subfigure shows the latency on SUN RGB-D. There are 8 stacked bar charts, and 2 neighboring bars are grouped together. As a result, there are 4 groups. The name of each group is CPU-CPU, GPU-EdgeTPU, GPU-CPU and GPU-EdgeTPU, from left to right. In each group, the left bar represents PointPainting(INT8) and right bar represents PointSplit(INT8).
  Every bar is three bars stacked. The bottom one is DeeplabV3+, the middle one is PointNet++ and the top one is Voting and Proposal.
  In the left bar in the first group(CPU-CPU), the bottom bar is 775ms, the middle bar is 1483ms, the top bar is 126ms.
  In the right bar in the first group(CPU-CPU), the bottom bar is 784ms, the middle bar is 1210ms, the top bar is 158ms.
  In the left bar in the second group(CPU-EdgeTPU), the bottom bar is 219ms, the middle bar is 3442ms, the top bar is 135ms.
  In the right bar in the second group(CPU-EdgeTPU), the bottom bar is 224ms, the middle bar is 2116ms, the top bar is 155ms.
  In the left bar in the third group(GPU-CPU), the bottom bar is 775ms, the middle bar is 1483ms, the top bar is 126ms.
  In the right bar in the third group(GPU-CPU), the bottom bar is 784ms, the middle bar is 1210ms, the top bar is 158ms.
  In the left bar in the fourth group(GPU-EdgeTPU), the bottom bar is 222ms, the middle bar is 631ms, the top bar is 73ms.
  In the right bar in the fourth group(GPU-EdgeTPU), the bottom bar is 227ms, the middle bar is 414ms, the top bar is 83ms.

  The second subfigure shows the latency on ScannetV2. There are 8 stacked bar charts, and 2 neighboring bars are grouped together. As a result, there are 4 groups. The name of each group is CPU-CPU, GPU-EdgeTPU, GPU-CPU and GPU-EdgeTPU, from left to right. In each group, the left bar represents PointPainting(INT8) and right bar represents PointSplit(INT8).
  Every bar is three bars stacked. The bottom one is DeeplabV3+, the middle one is PointNet++ and the top one is Voting and Proposal.
  In the left bar in the first group(CPU-CPU), the bottom bar is 1252ms, the middle bar is 8121ms, the top bar is 219ms.
  In the right bar in the first group(CPU-CPU), the bottom bar is 1100ms, the middle bar is 4303ms, the top bar is 234ms.
  In the left bar in the second group(CPU-EdgeTPU), the bottom bar is 373ms, the middle bar is 7312ms, the top bar is 160ms.
  In the right bar in the second group(CPU-EdgeTPU), the bottom bar is 353ms, the middle bar is 3955ms, the top bar is 171ms.
  In the left bar in the third group(GPU-CPU), the bottom bar is 1318ms, the middle bar is 2047ms, the top bar is 180ms.
  In the right bar in the third group(GPU-CPU), the bottom bar is 1038, the middle bar is 1509, the top bar is 163ms.
  In the left bar in the fourth group(GPU-EdgeTPU), the bottom bar is 332ms, the middle bar is 1056ms, the top bar is 80ms.
  In the right bar in the fourth group(GPU-EdgeTPU), the bottom bar is 327ms, the middle bar is 673ms, the top bar is 95ms.  
}
  \vspace{-2ex}
  \label{fig:Systembaseline}
\end{figure}

\vspace{1ex}\noindent
\textbf{More hardware configurations.} 
\rev{Next, Figure~\ref{fig:Systembaseline}  evaluates the impact of \proposal on various processor combinations in our platform: (1) CPU-CPU, (2) CPU-EdgeTPU, (3) GPU-CPU, and (4) GPU-EdgeTPU (i.e., our platform). In each combination, the first processor executes point manipulation (in PointNet++) and the second processor executes neural nets, such as PointNet (in PointNet++), Deeplabv3+ and voting/proposal modules (in VoteNet). 
The results show that hardware configuration significantly impacts latency. Specifically, using GPU as the first processor, instead of CPU, accelerates point manipulation, which reduces latency for running PointNet++. Using EdgeTPU as the second processor improves latency of all the neural nets compared to using CPU.
More importantly, the results verify that \proposal reduces latency on every hardware configuration compared to PointPainting (INT8). Specifically, \proposal improves latency performance most significantly in the CPU-CPU and CPU-EdgeTPU cases, 1.7$\times$ and 1.8$\times$, respectively. Note that PointPainting (INT8) provides significantly lower mAP than \proposal in Table~\ref{tab:pipeline}}.

\vspace{1ex}\noindent
\textbf{Layer-wise analysis.} 
\rev{To take a deeper look, Table~\ref{tab:latency_by_layer} shows per-layer latency of PointPainting (INT8) and PointNet++ (INT8) when using GPU and EdgeTPU without parallelization. As the layer proceeds, computation on GPU (point manipulation) monotonically decreases due to the smaller number of sampled points while that on EdgeTPU (PointNet) first increases and decreases again due to the trade-off between the input size and the number of channels.
The results verify that running point manipulation for SA-normal 
 on GPU while fusing 2D-3D information on EdgeTPU for SA-bias significantly reduces latency. In addition, given that GPU needs much more time than EdgeTPU at SA1, adding more layers to PointNet in SA1 and process the layers on EdgeTPU using the idle time might improve accuracy without sacrificing latency.
}

\input{tab/latency_by_layer.tex}

\input{tab/comm_overhead}
\vspace{1ex}\noindent
\textbf{Inter-processor communication.} 
Utilizing multiple accelerators requires data exchange between the accelerators, which brings a concern of inter-processor communication overhead. Table~\ref{tab:commOverhead} \rev{quantifies} the communication overhead by dividing \proposal's inference latency on a SUN RGB-D scene into communication and computation latency. To focus on PointNet++ operation, we exclude 2D-3D fusion.  
While latency on GPU is measured by NVIDIA profiler, without such a tool, that on EdgeTPU is estimated in the following way. We first measure the time required to execute each PointNet in EdgeTPU, denoted as $t_{p,total}$, which includes latency for both communication and computation: $t_{p,total} = t_{p,comp} + t_{p,comm}$. Then we build another PointNet that has the same size of input, output, and the number of parameters, but doubles the amount of computation. The time for executing the new PointNet model, $t_{p2,total}$, includes the same communication time but twice longer computation time: $t_{p2,total}= 2 \times t_{p,comp} + t_{p,comm}$. Therefore, we estimate computation latency on EdgeTPU as the difference between the two measurements: $t_{p,comp} = t_{p2,total}-t_{p,total}$. Then the remainder is regarded as communication latency: $t_{p,comm} = t_{p,total} - t_{p,comp}$.

The results verify that communication overhead on our platform is significant indeed, taking up 54.4\% of the total latency.
Specifically, communication time on EdgeTPU is 4.5$\times$ longer than that on GPU due to the use of a slower channel, PCIe Gen 2 x 1 (0.5 GB/s). 
With a short glimpse, the significant communication overhead seems to suggest that parallization among heterogeneous low-power processors might have limited gain. However, \textit{the real implication is opposite}: once resource-constrained hardware evolves further and solves the communication problem, which is actually happening these days, \proposal's inference speed can be nearly doubled.
%
With the latest off-the-shelf hardware equipped with multi-type accelerators, such as Apple's M1 architecture, we expect the field of on-device ML to evolve further with parallel processing.

%% file: tab/eval_sunrgbd_class_25.tex
\begin{table*}[t]
\centering
  \caption{Per-class accuracy (mAP) at IoU threshold 0.25 of various 3D object detectors on SUN RGB-D (our primary dataset). \proposal (FP32) provides the best accuracy for 4 out of 10 classes, resulting in the best overall mAP performance. After quantized, \proposal still performs comparably to PointPainting and significantly outperforms VoteNet.}
  \label{tab:evalSunrgbdClass_25}
\vspace{-3ex}
\resizebox{\linewidth}{!}{ 
  \begin{tabular}{ccccccccccccc}
    \toprule
    Item & Bathtub & Bed & Bookshelf & Chair & Desk & Dresser & Nightstand & Sofa & Table & Toilet & \vline & Overall\\
    \midrule
    VoteNet (FP32)                 & \textbf{72.4} & 84.0 & 25.3 & 74.1 & 24.2 & 30.0 & 61.4 & 61.6 & 49.7 & 86.8 & \vline & 56.9 \\
    PointPainting (FP32) & 68.0 & \textbf{86.5} & 29.6 & 74.1 & 24.6 & \textbf{39.9} & \textbf{61.8} & 77.9 & 49.3 & 90.0 & \vline & 60.2 \\
    
    RandomSplit (FP32)            & 61.9 & 85.6 & 33.8 & 74.5 & 26.4 & 38.7 & 61.7 & \textbf{79.7} & \textbf{52.8} & 88.9 & \vline & 60.4  \\
    \proposal (FP32)              & 69.0 & 86.0 & \textbf{34.0} & \textbf{74.9} & \textbf{27.0} & 39.7 & 60.1 & 78.5 & 51.8 & \textbf{92.5} & \vline & \textbf{61.4} \\
    \proposal (INT8)   & 62.7 & 86.3 & 33.0 & 74.4 & 25.5 & 39.3 & 58.9 & 77.6 & 50.6 & 90.5 & \vline & 59.9 \\
  \bottomrule

    \end{tabular}
} 
\vspace{-2ex}
\end{table*}

%% file: tab/pipeline.tex

\begin{table}[t]
\centering
\vspace{-1ex}
  \caption{\rev{mAP of various VoteNet-based 3D object detectors on SUN RGB-D and Scannet V2, measured at IoU thresholds of 0.25 and 0.5. \proposal provides the best mAP in most cases.}}
  \vspace{-3ex}
  \label{tab:pipeline}
\begin{tabular}{cccc}
\toprule
\multirow{3}{*}{Precision} & \multirow{2}{*}{Method}   & \multicolumn{2}{c}{Dataset}                  \\ 
    &   &   SUN RGB-D  & Scannet V2  \\
    &   &   @0.25 / @0.5   & @0.25 / @0.5  \\  \midrule
\multirow{4}{*}{FP32}       & VoteNet       & 56.9 / 31.1           & 54.9 / 30.4               \\
                            & PointPainting & 60.2 / \textbf{32.8}  & \textbf{56.4} / 31.7       \\
                            & RandomSplit   & 60.4 / 32.0           & 55.2 / 31.2             \\ 
                            & \proposal     & \textbf{61.4} / 32.7  & 56.1 / \textbf{32.4}  \\ \hline
\multirow{3}{*}{INT8}       & VoteNet       & 29.3 / 3.0             & 41.7 / 11.6             \\     
                            & PointPainting &  32.3 / 3.2            & 48.8 / 18.2             \\             
                            & \proposal     & \textbf{59.9} / \textbf{32.5} & \textbf{55.7} / \textbf{30.3}        \\ \bottomrule
\end{tabular}
\vspace{-3ex}
\end{table}

%% file: tab/groupfree3d.tex

\begin{table}[t]
\centering

  \caption{\rev{mAP of \proposal combined with GroupFree3D~\cite{liu2021group} and RepSurf~\cite{ran2022surface} on SUN RGB-D and Scannet V2. Scannet V2 experiments use 5 2D images for 2D-3D fusion. (6,256) means that the GroupFree3D model has 6 decoder layers and uses 256 object candidates.}}
  \vspace{-2ex}
  \label{tab:groupfree3d}
\resizebox{.9\linewidth}{!}{ 
    \begin{threeparttable}
    \begin{tabular}{ccc}
    \toprule
    \multirow{3}{*}{Method}   & \multicolumn{2}{c}{Dataset}                  \\ 
           &   SUN RGB-D  & Scannet V2  \\
           &   @0.25 / @0.5  & @0.25 / @0.5  \\ \midrule
    Baseline: GroupFree3D\textsuperscript{(6,256)}    & 58.0 / 38.3             & 63.7 / 38.8    \\
    Baseline + PointPainting  & 62.5 / \textbf{43.3}             & 66.7 / 41.2  \\
    Baseline + RandomSplit    & 61.9 / 40.4               & 66.6 / 33.7  \\
    Baseline + \proposal       & \textbf{62.6} / 42.5    & \textbf{67.8} / \textbf{45.4} \\ 
    \hline
    Baseline: RepSurf-U + GroupFree3D\textsuperscript{(6,256)}     & 61.4 / 41.3             & 65.0 / 41.0    \\
    Baseline + PointPainting  & 63.1 / 41.8             & 67.4 / 42.7  \\
    Baseline + RandomSplit    & 62.5 / 40.8               & 67.0 / 43.3  \\
    Baseline + \proposal       & \textbf{63.5} / \textbf{42.1}    & \textbf{68.5} / \textbf{46.7} \\
    \bottomrule
    
    \end{tabular}    
    \end{threeparttable}    
}

\vspace{-2ex}

\end{table}

%% file: tab/bfps_sa.tex
\begin{table}[t]  
    \centering
    \caption{Accuracy of \proposal on SUN RGB-D with varying $w_0$ for the semantics-aware biased point sampling. The performance is maximized when point sampling is slightly biased toward foreground points.}
    \vspace{-2ex}
    \label{tab:bfps_weights}
    \begin{tabular}{ccccccc}
        \toprule    
        Weight & 0.5 & 1.0 & 1.5 & 2.0 & 2.5 & 3.5 \\
        \midrule
        mAP & 60.3 & 60.4 & 61.3 & \textbf{61.4} & 59.6 & 59.4 \\            
        \bottomrule
    \end{tabular}
\vspace{-2ex}
\end{table}

%% file: tab/quant_error.tex
\begin{table*}[t]
  \centering  
  \caption{mAP at IoU threshold 0.25 of \proposal on SUN RGB-D and Scannet V2, with various quantization methods. While  performing similarly to the most fine-grained channel-wise method, our role-based group-wise quantization method remarkably reduces the number of quantization parameters, 67$\times$ and 71$\times$ less parameters on SUN RGB-D and Scannet V2, respectively.}
  \vspace{-2.5ex}
  \label{tab:quantError}
  \resizebox{\linewidth}{!}{ 
  \begin{tabular}{cc|ccr|ccr}
    \toprule
    \multirow{2}{*}{Quant. method} & \multirow{2}{*}{Precision} & \multicolumn{3}{c|}{SUN RGB-D} &  \multicolumn{3}{c}{ScannetV2} \\
     & & mAP & Quant. error & \begin{tabular}[c]{@{}c@{}} \# of quant.\\ parameters\end{tabular} & mAP & Quant. error & \begin{tabular}[c]{@{}c@{}}\# of quant.\\ parameters\end{tabular}  \\
    \midrule
    No quant. & FP32 & 61.4 & - & - & 56.1 & - & -\\ \hline
    Layer-wise & INT8& 24.2 & 37.2 & 8 & 51.9 & 4.2 & 8 \\
    Group-wise & INT8 & 26.3 & 35.1 & 20 & 52.3 & 3.8 & 20 \\
    Channel-wise & INT8 & 61.0 & 0.4 & 1352 & 55.5 & 0.6 & 1424 \\
    \textbf{Role-based group-wise (ours)} & INT8 & 59.9 & 1.5 & 20 & 55.4 & 0.7 & 20 \\
  \bottomrule
\end{tabular}
} 
\vspace{-2ex}
\end{table*}

%% file: tab/latency_by_layer.tex
\begin{table}[t]
\centering
  \caption{\rev{Per-layer latency of PointPainting (INT8) and PointNet++ (INT8) with sequential pipelining.}}
  \vspace{-3ex}
  \label{tab:latency_by_layer}
\begin{tabular}{ccc}
\toprule
Layers & GPU  & EdgeTPU   \\ 

\midrule
2D-3D fusion     & -        & 222 ms         \\
SA1              & 199 ms   & 47 ms       \\
SA2              & 52  ms   & 71 ms       \\
SA3              & 25  ms   & 84 ms       \\
SA4              & 20  ms   & 21 ms       \\ 
\bottomrule
\end{tabular}
\vspace{-3ex}
\end{table}

%% file: tab/comm_overhead.tex
\begin{table}[t]
\centering
  \caption{Latency in communication and computation on GPU and EdgeTPU  when \proposal processes a single SUN RGB-D scene. For the ease of measurement, the latency to run DeeplabV3+ is not included and multithreading is not used (SA-normal and SA-bias are executed sequentially).}
  \label{tab:commOverhead}
    \vspace{-3ex}
  \begin{tabular}{cccc}
    \toprule
    
    \multirow{2}{*}{Processor}   & \multicolumn{3}{c}{Latency (ms)} \\
                                & Communication & Computation & Total \\
    \midrule
    GPU                     & 80 & 248 & 328   \\
    EdgeTPU (estimates)     & 360 & 121 & 481   \\
  \bottomrule

    \end{tabular}
    \vspace{-3ex}
\end{table}

%% file: sec/7_discussion.tex
\vspace{-1.5ex}
\section{Discussion}

\rev{In this section, we discuss applicability and limitations of \proposal. In addition, we present practical challenges (i.e., entry barriers) for researching on-device 3D object detection, which we have experienced while developing \proposal, the first framework for 100\% on-device 3D object detection with heterogeneous accelerators.}

\vspace{-1ex}
\subsection{Generalization and Limitations}


\noindent\textbf{2D semantics-aware biased point sampling (\S{\ref{sec:ps_biased})}).} 
\rev{Although our biased point sampling method is implemented on PointNet++, the idea of biased point sampling is not specific to PointNet++. Our method can be directly applied to any DNN that utilizes farthest point sampling (FPS) and easily adapted for other point sampling techniques. A point sampling technique has its own metric (e.g., distance or density) and our technique is applied to the sampling method by slightly modifying the metric with point semantics. 
For example, in case of a density-based sampling technique~\cite{liu2020morphing}, we can simply boost a point’s density-based metric value if the point is in a specific group that needs to be sampled more intensely.} 
\rev{On the other hand, there are 3D object detection networks that do not exploit point sampling (i.e., voxel-based 3D object detectors~\cite{deng2021voxel, mao2021voxel}) where our point sampling technique is not applicable.}

\vspace{1ex}
\noindent\textbf{PointNet++ parallelization (\S{\ref{sec:pointnet_opt}}).}
\rev{It is important that our parallelization technique is not specific to VoteNet nor GroupFree3D but their backbone PointNet++, which is one of the most widely used 3D backbone networks. Many recent state-of-the-art  models for 3D object detection on SUN RGB-D (our primary dataset) and ScannetV2 (our secondary dataset) employ PointNet++ as their backbone~\cite{yang2022boosting, wang2022multimodal,ran2022surface,wang2022rbgnet,liu2021group,cheng2021back,zhang2020h3dnet,misra2021end}. Specifically, out of top 10 ranked methods,  7 methods on SUN RGB-D and 6 methods on ScannetV2 use PointNet++, showing that PointNet++-based models are dominating currently.}

\vspace{1ex}
\noindent\textbf{Role-based group-wise quantization (\S{\ref{sec:role-quant}}).} 
\rev{The role-based group-wise quantization is motivated by our observation that a layer's weight/activation distributions are impacted by what role each node has. Therefore, the role-based grouping scheme can be applied to any network layer that has multiple roles, not only for VoteNet. It would be an interesting future work to evaluate the impact of role-based grouping on other 3D object detectors. In addition, although we focus on quantization in this work, investigating other compression approaches, such as knowledge distillation and pruning, can be valuable future work.}

\vspace{-1ex}
\subsection{Challenges in On-device 3D Object Detection}

\rev{The field of 3D object detection has experienced significant growth in recent years within the deep learning community, with a range of datasets and model implementations now available. However, the deployment of state-of-the-art models on resource-constrained devices presents several nontrivial challenges.}

\vspace{1ex}
\noindent\textbf{Model size and complexity.} 
\rev{
Given that popular leaderboards on 3D object detection primarily evaluate accuracy, many of the top-ranked models rely on transformers~\cite{yang2022boosting, wang2022multimodal,ran2022surface,liu2021group} or custom modules~\cite{yang2022boosting,wang2022cagroup3d,rukhovich2022fcaf3d,wang2022rbgnet,chen2020hierarchical,cheng2021back,vu2022softgroup} that are too heavy to run on resource-constrained devices. 
For example, DeMF~\cite{yang2022boosting}, which currently ranks first on the SUN RGB-D leaderboard, reaches a peak GPU memory of 2.5 GB and requires 173 GFLOPS for its 2D detector with deformable attention~\cite{zhu2020deformable}. Effective model compression techniques must therefore be developed in order to enable the deployment of the latest models on edge devices.}

\vspace{1ex}
\noindent\textbf{Implementation burden.}
\rev{Although the latest 3D object detectors are implemented using Pytorch, which consumes significant resources, they are not currently implemented on  edge-friendly platforms, such as TensorFlow Lite and MNN~\cite{jiang2020mnn}.  
Furthermore, many state-of-the-art models rely on recent software packages, such as \texttt{mmdetection3d} and \texttt{Minkowski}, which are not currently supported by edge devices. As a result, significant time and effort is required to re-implement state-of-the-art models on lightweight platforms that perform comparably to their Pytorch versions. 
We believe that our open implementation of VoteNet on TensorFlow Lite can accelerate future research on on-device 3D object detection.}


%% file: sec/8_conclusion.tex
\vspace{-1.5ex}
\section{Conclusion}

This work began when we observed the emergence of multi-type low-power accelerators with different pros and cons. We envisioned that in the era of multi-type accelerators, a new class of intelligent tasks that used to be too heavy can be viable in the on-device ML regime when these accelerators are utilized synergistically.
To investigate the potential, we have built a low-power hardware platform including  both GPU and NPU, and studied on-device 3D object detection with 2D-3D information fusion.

Specifically, we propose \proposal, a novel 3D object detection framework that provides system-algorithm joint optimization. 
First, \proposal catches the difference between point manipulation and neural net operation in a representative 3D feature extractor (PoinNet++), executing the former on GPU and the latter on NPU. 
Second, \proposal creates two separate but synergistic feature extraction pipelines by augmenting a point cloud scene with 2D semantic information (i.e., semantics-aware biased point sampling). The PointNet++ structure is further optimized to maximize accuracy and efficiency in the \proposal framework.
Third, \proposal contains role-based group-wise quantization that quantizes a 3D object detector with a small number of parameters without sacrificing accuracy. 
Our experiments demonstrate the effectiveness of \proposal in terms of both accuracy and latency. 
We believe that this work, by showing the potential of recently available edge devices equipped with heterogeneous low-power processors, and by providing open implementation, can inspire other researchers to run more various complex tasks on the new class of edge devices. 
\vspace{-1.5ex}

%% file: main.bbl

\begin{thebibliography}{74}


\ifx \showCODEN    \undefined \def \showCODEN     #1{\unskip}     \fi
\ifx \showDOI      \undefined \def \showDOI       #1{#1}\fi
\ifx \showISBNx    \undefined \def \showISBNx     #1{\unskip}     \fi
\ifx \showISBNxiii \undefined \def \showISBNxiii  #1{\unskip}     \fi
\ifx \showISSN     \undefined \def \showISSN      #1{\unskip}     \fi
\ifx \showLCCN     \undefined \def \showLCCN      #1{\unskip}     \fi
\ifx \shownote     \undefined \def \shownote      #1{#1}          \fi
\ifx \showarticletitle \undefined \def \showarticletitle #1{#1}   \fi
\ifx \showURL      \undefined \def \showURL       {\relax}        \fi
\providecommand\bibfield[2]{#2}
\providecommand\bibinfo[2]{#2}
\providecommand\natexlab[1]{#1}
\providecommand\showeprint[2][]{arXiv:#2}

\bibitem[Apicharttrisorn et~al\mbox{.}(2019)]%
        {apicharttrisorn2019frugal}
\bibfield{author}{\bibinfo{person}{Kittipat Apicharttrisorn} {et~al\mbox{.}}}
  \bibinfo{year}{2019}\natexlab{}.
\newblock \showarticletitle{Frugal following: Power thrifty object detection
  and tracking for mobile augmented reality}. In
  \bibinfo{booktitle}{\emph{Proceedings of the 17th ACM Conference on Embedded
  Networked Sensor Systems}}. \bibinfo{pages}{96--109}.
\newblock


\bibitem[Cai et~al\mbox{.}(2021)]%
        {Cai_Li_Yuan_Niu_Li_Tang_Ren_Wang_2021}
\bibfield{author}{\bibinfo{person}{Yuxuan Cai} {et~al\mbox{.}}}
  \bibinfo{year}{2021}\natexlab{}.
\newblock \showarticletitle{YOLObile: Real-Time Object Detection on Mobile
  Devices via Compression-Compilation Co-Design}.
\newblock \bibinfo{journal}{\emph{Proceedings of the AAAI Conference on
  Artificial Intelligence}} \bibinfo{volume}{35}, \bibinfo{number}{2}
  (\bibinfo{date}{May} \bibinfo{year}{2021}), \bibinfo{pages}{955--963}.
\newblock


\bibitem[Charles et~al\mbox{.}(2017)]%
        {8099499}
\bibfield{author}{\bibinfo{person}{Qi. Charles} {et~al\mbox{.}}}
  \bibinfo{year}{2017}\natexlab{}.
\newblock \showarticletitle{PointNet: Deep Learning on Point Sets for 3D
  Classification and Segmentation}. In \bibinfo{booktitle}{\emph{Proceedings of
  the IEEE/CVF Conference on Computer Vision and Pattern Recognition (CVPR)}}.
  \bibinfo{pages}{77--85}.
\newblock
\showISSN{1063-6919}


\bibitem[Chen et~al\mbox{.}(2020a)]%
        {Chen_2020_CVPR}
\bibfield{author}{\bibinfo{person}{Jintai Chen} {et~al\mbox{.}}}
  \bibinfo{year}{2020}\natexlab{a}.
\newblock \showarticletitle{A Hierarchical Graph Network for 3D Object
  Detection on Point Clouds}. In \bibinfo{booktitle}{\emph{Proceedings of the
  IEEE/CVF Conference on Computer Vision and Pattern Recognition (CVPR)}}.
\newblock


\bibitem[Chen et~al\mbox{.}(2020b)]%
        {chen2020hierarchical}
\bibfield{author}{\bibinfo{person}{Jintai Chen} {et~al\mbox{.}}}
  \bibinfo{year}{2020}\natexlab{b}.
\newblock \showarticletitle{A hierarchical graph network for 3D object
  detection on point clouds}. In \bibinfo{booktitle}{\emph{Proceedings of the
  IEEE/CVF Conference on Computer Vision and Pattern Recognition (CVPR)}}.
  \bibinfo{pages}{392--401}.
\newblock


\bibitem[Chen et~al\mbox{.}(2018a)]%
        {chen2018marvel}
\bibfield{author}{\bibinfo{person}{Kaifei Chen} {et~al\mbox{.}}}
  \bibinfo{year}{2018}\natexlab{a}.
\newblock \showarticletitle{Marvel: Enabling mobile augmented reality with low
  energy and low latency}. In \bibinfo{booktitle}{\emph{Proceedings of the 16th
  ACM Conference on Embedded Networked Sensor Systems}}.
  \bibinfo{pages}{292--304}.
\newblock


\bibitem[Chen et~al\mbox{.}(2014)]%
        {chen2014semantic}
\bibfield{author}{\bibinfo{person}{Liang-Chieh Chen} {et~al\mbox{.}}}
  \bibinfo{year}{2014}\natexlab{}.
\newblock \showarticletitle{Semantic image segmentation with deep convolutional
  nets and fully connected crfs}.
\newblock \bibinfo{journal}{\emph{arXiv preprint arXiv:1412.7062}}
  (\bibinfo{year}{2014}).
\newblock


\bibitem[Chen et~al\mbox{.}(2017a)]%
        {chen2017deeplab}
\bibfield{author}{\bibinfo{person}{Liang-Chieh Chen} {et~al\mbox{.}}}
  \bibinfo{year}{2017}\natexlab{a}.
\newblock \showarticletitle{Deeplab: Semantic image segmentation with deep
  convolutional nets, atrous convolution, and fully connected crfs}.
\newblock \bibinfo{journal}{\emph{IEEE transactions on pattern analysis and
  machine intelligence}} \bibinfo{volume}{40}, \bibinfo{number}{4}
  (\bibinfo{year}{2017}), \bibinfo{pages}{834--848}.
\newblock


\bibitem[Chen et~al\mbox{.}(2017b)]%
        {chen2017rethinking}
\bibfield{author}{\bibinfo{person}{Liang-Chieh Chen} {et~al\mbox{.}}}
  \bibinfo{year}{2017}\natexlab{b}.
\newblock \showarticletitle{Rethinking atrous convolution for semantic image
  segmentation}.
\newblock \bibinfo{journal}{\emph{arXiv preprint arXiv:1706.05587}}
  (\bibinfo{year}{2017}).
\newblock


\bibitem[Chen et~al\mbox{.}(2018b)]%
        {chen2018encoder}
\bibfield{author}{\bibinfo{person}{Liang-Chieh Chen} {et~al\mbox{.}}}
  \bibinfo{year}{2018}\natexlab{b}.
\newblock \showarticletitle{Encoder-decoder with atrous separable convolution
  for semantic image segmentation}. In \bibinfo{booktitle}{\emph{Proceedings of
  the European conference on computer vision (ECCV)}}.
  \bibinfo{pages}{801--818}.
\newblock


\bibitem[Chen et~al\mbox{.}(2021)]%
        {Chen_2021_CVPR}
\bibfield{author}{\bibinfo{person}{Peng Chen} {et~al\mbox{.}}}
  \bibinfo{year}{2021}\natexlab{}.
\newblock \showarticletitle{AQD: Towards Accurate Quantized Object Detection}.
  In \bibinfo{booktitle}{\emph{Proceedings of the IEEE/CVF Conference on
  Computer Vision and Pattern Recognition (CVPR)}}. \bibinfo{pages}{104--113}.
\newblock


\bibitem[Chen et~al\mbox{.}(2017c)]%
        {Chen_2017_CVPR}
\bibfield{author}{\bibinfo{person}{Xiaozhi Chen} {et~al\mbox{.}}}
  \bibinfo{year}{2017}\natexlab{c}.
\newblock \showarticletitle{Multi-View 3D Object Detection Network for
  Autonomous Driving}. In \bibinfo{booktitle}{\emph{Proceedings of the IEEE
  Conference on Computer Vision and Pattern Recognition (CVPR)}}.
\newblock


\bibitem[Cheng et~al\mbox{.}(2021)]%
        {cheng2021back}
\bibfield{author}{\bibinfo{person}{Bowen Cheng} {et~al\mbox{.}}}
  \bibinfo{year}{2021}\natexlab{}.
\newblock \showarticletitle{Back-tracing representative points for voting-based
  3d object detection in point clouds}. In
  \bibinfo{booktitle}{\emph{Proceedings of the IEEE/CVF Conference on Computer
  Vision and Pattern Recognition (CVPR)}}. \bibinfo{pages}{8963--8972}.
\newblock


\bibitem[Choi et~al\mbox{.}(2022)]%
        {choi2022scriptpainter}
\bibfield{author}{\bibinfo{person}{Yousung Choi} {et~al\mbox{.}}}
  \bibinfo{year}{2022}\natexlab{}.
\newblock \showarticletitle{ScriptPainter: Vision-based, On-device Test Script
  Generation for Mobile Systems}. In \bibinfo{booktitle}{\emph{2022 21st
  ACM/IEEE IPSN}}. IEEE, \bibinfo{pages}{477--490}.
\newblock


\bibitem[Dai et~al\mbox{.}(2017)]%
        {dai2017scannet}
\bibfield{author}{\bibinfo{person}{Angela Dai} {et~al\mbox{.}}}
  \bibinfo{year}{2017}\natexlab{}.
\newblock \showarticletitle{Scannet: Richly-annotated 3d reconstructions of
  indoor scenes}. In \bibinfo{booktitle}{\emph{Proceedings of the IEEE/CVF
  Conference on Computer Vision and Pattern Recognition (CVPR)}}.
  \bibinfo{pages}{5828--5839}.
\newblock


\bibitem[Deng et~al\mbox{.}(2021)]%
        {deng2021voxel}
\bibfield{author}{\bibinfo{person}{Jiajun Deng} {et~al\mbox{.}}}
  \bibinfo{year}{2021}\natexlab{}.
\newblock \showarticletitle{Voxel r-cnn: Towards high performance voxel-based
  3d object detection}. In \bibinfo{booktitle}{\emph{Proceedings of the AAAI
  Conference on Artificial Intelligence}}, Vol.~\bibinfo{volume}{35}.
  \bibinfo{pages}{1201--1209}.
\newblock


\bibitem[Ding et~al\mbox{.}(2019)]%
        {ding2019req}
\bibfield{author}{\bibinfo{person}{Caiwen Ding} {et~al\mbox{.}}}
  \bibinfo{year}{2019}\natexlab{}.
\newblock \showarticletitle{REQ-YOLO: A resource-aware, efficient quantization
  framework for object detection on FPGAs}. In
  \bibinfo{booktitle}{\emph{Proceedings of the 2019 ACM/SIGDA International
  Symposium on Field-Programmable Gate Arrays}}. \bibinfo{pages}{33--42}.
\newblock


\bibitem[Drozdzal et~al\mbox{.}(2016)]%
        {drozdzal2016importance}
\bibfield{author}{\bibinfo{person}{Michal Drozdzal} {et~al\mbox{.}}}
  \bibinfo{year}{2016}\natexlab{}.
\newblock \showarticletitle{The importance of skip connections in biomedical
  image segmentation}.
\newblock In \bibinfo{booktitle}{\emph{Deep learning and data labeling for
  medical applications}}. \bibinfo{publisher}{Springer},
  \bibinfo{pages}{179--187}.
\newblock


\bibitem[Frankle et~al\mbox{.}(2019)]%
        {DBLP:conf/iclr/FrankleC19}
\bibfield{author}{\bibinfo{person}{Jonathan Frankle} {et~al\mbox{.}}}
  \bibinfo{year}{2019}\natexlab{}.
\newblock \showarticletitle{The Lottery Ticket Hypothesis: Finding Sparse,
  Trainable Neural Networks}. In \bibinfo{booktitle}{\emph{7th International
  Conference on Learning Representations, (ICLR)}}.
\newblock


\bibitem[Guan et~al\mbox{.}(2022)]%
        {guan2022deepmix}
\bibfield{author}{\bibinfo{person}{Yongjie Guan} {et~al\mbox{.}}}
  \bibinfo{year}{2022}\natexlab{}.
\newblock \showarticletitle{DeepMix: mobility-aware, lightweight, and hybrid 3D
  object detection for headsets}. In \bibinfo{booktitle}{\emph{Proceedings of
  the 20th Annual International Conference on Mobile Systems, Applications and
  Services}}. \bibinfo{pages}{28--41}.
\newblock


\bibitem[Han et~al\mbox{.}(2015)]%
        {10.5555/2969239.2969366}
\bibfield{author}{\bibinfo{person}{Song Han} {et~al\mbox{.}}}
  \bibinfo{year}{2015}\natexlab{}.
\newblock \showarticletitle{Learning Both Weights and Connections for Efficient
  Neural Networks}. In \bibinfo{booktitle}{\emph{Proceedings of the 28th
  International Conference on Neural Information Processing Systems - Volume
  1}}. \bibinfo{pages}{1135–1143}.
\newblock


\bibitem[Han et~al\mbox{.}(2016)]%
        {DBLP:journals/corr/HanMD15}
\bibfield{author}{\bibinfo{person}{Song Han} {et~al\mbox{.}}}
  \bibinfo{year}{2016}\natexlab{}.
\newblock \showarticletitle{Deep Compression: Compressing Deep Neural Network
  with Pruning, Trained Quantization and Huffman Coding}. In
  \bibinfo{booktitle}{\emph{4th International Conference on Learning
  Representations, (ICLR)}}.
\newblock


\bibitem[Hinton et~al\mbox{.}(2015)]%
        {44873}
\bibfield{author}{\bibinfo{person}{Geoffrey Hinton} {et~al\mbox{.}}}
  \bibinfo{year}{2015}\natexlab{}.
\newblock \showarticletitle{Distilling the Knowledge in a Neural Network}. In
  \bibinfo{booktitle}{\emph{NIPS Deep Learning and Representation Learning
  Workshop}}.
\newblock


\bibitem[Hou et~al\mbox{.}(2019)]%
        {hou20193d}
\bibfield{author}{\bibinfo{person}{Ji Hou} {et~al\mbox{.}}}
  \bibinfo{year}{2019}\natexlab{}.
\newblock \showarticletitle{3d-sis: 3d semantic instance segmentation of rgb-d
  scans}. In \bibinfo{booktitle}{\emph{Proceedings of the IEEE/CVF Conference
  on Computer Vision and Pattern Recognition (CVPR)}}.
  \bibinfo{pages}{4421--4430}.
\newblock


\bibitem[Howard et~al\mbox{.}(2019)]%
        {Howard_2019_ICCV}
\bibfield{author}{\bibinfo{person}{Andrew Howard} {et~al\mbox{.}}}
  \bibinfo{year}{2019}\natexlab{}.
\newblock \showarticletitle{Searching for MobileNetV3}. In
  \bibinfo{booktitle}{\emph{Proceedings of the IEEE/CVF International
  Conference on Computer Vision (ICCV)}}.
\newblock


\bibitem[Huang et~al\mbox{.}(2021)]%
        {huang2021codenet}
\bibfield{author}{\bibinfo{person}{Qijing Huang} {et~al\mbox{.}}}
  \bibinfo{year}{2021}\natexlab{}.
\newblock \showarticletitle{Codenet: Efficient deployment of input-adaptive
  object detection on embedded fpgas}. In \bibinfo{booktitle}{\emph{The 2021
  ACM/SIGDA International Symposium on Field-Programmable Gate Arrays}}.
  \bibinfo{pages}{206--216}.
\newblock


\bibitem[Hubara et~al\mbox{.}(2016)]%
        {hubara2016binarized}
\bibfield{author}{\bibinfo{person}{Itay Hubara} {et~al\mbox{.}}}
  \bibinfo{year}{2016}\natexlab{}.
\newblock \showarticletitle{Binarized neural networks}.
\newblock \bibinfo{journal}{\emph{Advances in neural information processing
  systems}}  \bibinfo{volume}{29} (\bibinfo{year}{2016}).
\newblock


\bibitem[Jacob et~al\mbox{.}(2018)]%
        {Jacob_2018_CVPR}
\bibfield{author}{\bibinfo{person}{Benoit Jacob} {et~al\mbox{.}}}
  \bibinfo{year}{2018}\natexlab{}.
\newblock \showarticletitle{Quantization and Training of Neural Networks for
  Efficient Integer-Arithmetic-Only Inference}. In
  \bibinfo{booktitle}{\emph{Proceedings of the IEEE/CVF Conference on Computer
  Vision and Pattern Recognition (CVPR)}}.
\newblock


\bibitem[Jiang et~al\mbox{.}(2020)]%
        {jiang2020mnn}
\bibfield{author}{\bibinfo{person}{Xiaotang Jiang} {et~al\mbox{.}}}
  \bibinfo{year}{2020}\natexlab{}.
\newblock \showarticletitle{Mnn: A universal and efficient inference engine}.
\newblock \bibinfo{journal}{\emph{Proceedings of Machine Learning and Systems}}
   \bibinfo{volume}{2} (\bibinfo{year}{2020}), \bibinfo{pages}{1--13}.
\newblock


\bibitem[Kim et~al\mbox{.}(2020)]%
        {9262933}
\bibfield{author}{\bibinfo{person}{Bogil Kim} {et~al\mbox{.}}}
  \bibinfo{year}{2020}\natexlab{}.
\newblock \showarticletitle{Energy-Efficient Acceleration of Deep Neural
  Networks on Realtime-Constrained Embedded Edge Devices}.
\newblock \bibinfo{journal}{\emph{IEEE Access}}  \bibinfo{volume}{8}
  (\bibinfo{year}{2020}), \bibinfo{pages}{216259--216270}.
\newblock


\bibitem[Kisantal et~al\mbox{.}(2019)]%
        {Kisantal_2019}
\bibfield{author}{\bibinfo{person}{Mate Kisantal} {et~al\mbox{.}}}
  \bibinfo{year}{2019}\natexlab{}.
\newblock \showarticletitle{Augmentation for small object detection}. In
  \bibinfo{booktitle}{\emph{9th International Conference on Advances in
  Computing and Information Technology ({ACITY} 2019)}}.
  \bibinfo{publisher}{Aircc Publishing Corporation}.
\newblock


\bibitem[Krishnamoorthi(2018)]%
        {krishnamoorthi2018quantizing}
\bibfield{author}{\bibinfo{person}{Raghuraman Krishnamoorthi}.}
  \bibinfo{year}{2018}\natexlab{}.
\newblock \showarticletitle{Quantizing deep convolutional networks for
  efficient inference: A whitepaper}.
\newblock \bibinfo{journal}{\emph{arXiv preprint arXiv:1806.08342}}
  (\bibinfo{year}{2018}).
\newblock


\bibitem[Li et~al\mbox{.}(2016)]%
        {li2016ternary}
\bibfield{author}{\bibinfo{person}{Fengfu Li} {et~al\mbox{.}}}
  \bibinfo{year}{2016}\natexlab{}.
\newblock \showarticletitle{Ternary weight networks}.
\newblock \bibinfo{journal}{\emph{arXiv preprint arXiv:1605.04711}}
  (\bibinfo{year}{2016}).
\newblock


\bibitem[Li et~al\mbox{.}(2019)]%
        {Li_2019_CVPR}
\bibfield{author}{\bibinfo{person}{Rundong Li} {et~al\mbox{.}}}
  \bibinfo{year}{2019}\natexlab{}.
\newblock \showarticletitle{Fully Quantized Network for Object Detection}. In
  \bibinfo{booktitle}{\emph{Proceedings of the IEEE/CVF Conference on Computer
  Vision and Pattern Recognition (CVPR)}}.
\newblock


\bibitem[Lin et~al\mbox{.}(2014)]%
        {lin2014microsoft}
\bibfield{author}{\bibinfo{person}{Tsung-Yi Lin} {et~al\mbox{.}}}
  \bibinfo{year}{2014}\natexlab{}.
\newblock \showarticletitle{Microsoft coco: Common objects in context}. In
  \bibinfo{booktitle}{\emph{Proceedings of the European conference on computer
  vision (ECCV)}}. Springer, \bibinfo{pages}{740--755}.
\newblock


\bibitem[Liu et~al\mbox{.}(2020)]%
        {liu2020morphing}
\bibfield{author}{\bibinfo{person}{Minghua Liu} {et~al\mbox{.}}}
  \bibinfo{year}{2020}\natexlab{}.
\newblock \showarticletitle{Morphing and sampling network for dense point cloud
  completion}. In \bibinfo{booktitle}{\emph{Proceedings of the AAAI Conference
  on Artificial Intelligence}}, Vol.~\bibinfo{volume}{34}.
  \bibinfo{pages}{11596--11603}.
\newblock


\bibitem[Liu et~al\mbox{.}(2019)]%
        {liu2019point}
\bibfield{author}{\bibinfo{person}{Zhijian Liu} {et~al\mbox{.}}}
  \bibinfo{year}{2019}\natexlab{}.
\newblock \showarticletitle{Point-voxel cnn for efficient 3d deep learning}.
\newblock \bibinfo{journal}{\emph{arXiv preprint arXiv:1907.03739}}
  (\bibinfo{year}{2019}).
\newblock


\bibitem[Liu et~al\mbox{.}(2021)]%
        {liu2021group}
\bibfield{author}{\bibinfo{person}{Ze Liu} {et~al\mbox{.}}}
  \bibinfo{year}{2021}\natexlab{}.
\newblock \showarticletitle{Group-free 3d object detection via transformers}.
  In \bibinfo{booktitle}{\emph{Proceedings of the IEEE/CVF International
  Conference on Computer Vision (ICCV)}}. \bibinfo{pages}{2949--2958}.
\newblock


\bibitem[Long et~al\mbox{.}(2015)]%
        {long2015fully}
\bibfield{author}{\bibinfo{person}{Jonathan Long} {et~al\mbox{.}}}
  \bibinfo{year}{2015}\natexlab{}.
\newblock \showarticletitle{Fully convolutional networks for semantic
  segmentation}. In \bibinfo{booktitle}{\emph{Proceedings of the IEEE
  conference on computer vision and pattern recognition}}.
  \bibinfo{pages}{3431--3440}.
\newblock


\bibitem[Mao et~al\mbox{.}(2021)]%
        {mao2021voxel}
\bibfield{author}{\bibinfo{person}{Jiageng Mao} {et~al\mbox{.}}}
  \bibinfo{year}{2021}\natexlab{}.
\newblock \showarticletitle{Voxel transformer for 3d object detection}. In
  \bibinfo{booktitle}{\emph{Proceedings of the IEEE/CVF International
  Conference on Computer Vision (ICCV)}}. \bibinfo{pages}{3164--3173}.
\newblock


\bibitem[Maturana et~al\mbox{.}(2015)]%
        {7353481}
\bibfield{author}{\bibinfo{person}{Daniel Maturana} {et~al\mbox{.}}}
  \bibinfo{year}{2015}\natexlab{}.
\newblock \showarticletitle{VoxNet: A 3D Convolutional Neural Network for
  real-time object recognition}. In \bibinfo{booktitle}{\emph{2015 IEEE/RSJ
  International Conference on Intelligent Robots and Systems (IROS)}}.
  \bibinfo{pages}{922--928}.
\newblock


\bibitem[Misra et~al\mbox{.}(2021)]%
        {misra2021end}
\bibfield{author}{\bibinfo{person}{Ishan Misra} {et~al\mbox{.}}}
  \bibinfo{year}{2021}\natexlab{}.
\newblock \showarticletitle{An end-to-end transformer model for 3d object
  detection}. In \bibinfo{booktitle}{\emph{Proceedings of the IEEE/CVF
  International Conference on Computer Vision (ICCV)}}.
  \bibinfo{pages}{2906--2917}.
\newblock


\bibitem[Qi et~al\mbox{.}(2020)]%
        {qi2020imvotenet}
\bibfield{author}{\bibinfo{person}{Charles Qi} {et~al\mbox{.}}}
  \bibinfo{year}{2020}\natexlab{}.
\newblock \showarticletitle{Imvotenet: Boosting 3d object detection in point
  clouds with image votes}. In \bibinfo{booktitle}{\emph{Proceedings of the
  IEEE/CVF Conference on Computer Vision and Pattern Recognition (CVPR)}}.
\newblock


\bibitem[Qi et~al\mbox{.}(2017)]%
        {qi2017pointnet++}
\bibfield{author}{\bibinfo{person}{Charles~R Qi} {et~al\mbox{.}}}
  \bibinfo{year}{2017}\natexlab{}.
\newblock \showarticletitle{Pointnet++: Deep hierarchical feature learning on
  point sets in a metric space}.
\newblock \bibinfo{journal}{\emph{arXiv preprint arXiv:1706.02413}}
  (\bibinfo{year}{2017}).
\newblock


\bibitem[Qi et~al\mbox{.}(2018)]%
        {qi2018frustum}
\bibfield{author}{\bibinfo{person}{Charles~R Qi} {et~al\mbox{.}}}
  \bibinfo{year}{2018}\natexlab{}.
\newblock \showarticletitle{Frustum pointnets for 3d object detection from
  rgb-d data}. In \bibinfo{booktitle}{\emph{Proceedings of the IEEE/CVF
  Conference on Computer Vision and Pattern Recognition (CVPR)}}.
  \bibinfo{pages}{918--927}.
\newblock


\bibitem[Qi et~al\mbox{.}(2019)]%
        {qi2019deep}
\bibfield{author}{\bibinfo{person}{Charles~R Qi} {et~al\mbox{.}}}
  \bibinfo{year}{2019}\natexlab{}.
\newblock \showarticletitle{Deep hough voting for 3d object detection in point
  clouds}. In \bibinfo{booktitle}{\emph{Proceedings of the IEEE/CVF
  International Conference on Computer Vision (ICCV)}}.
  \bibinfo{pages}{9277--9286}.
\newblock


\bibitem[Qian et~al\mbox{.}(2020)]%
        {xie2020mlcvnet}
\bibfield{author}{\bibinfo{person}{Xie Qian} {et~al\mbox{.}}}
  \bibinfo{year}{2020}\natexlab{}.
\newblock \showarticletitle{MLCVNet: Multi-Level Context VoteNet for 3D Object
  Detection}. In \bibinfo{booktitle}{\emph{Proceedings of the IEEE/CVF
  Conference on Computer Vision and Pattern Recognition (CVPR)}}.
\newblock


\bibitem[Ran et~al\mbox{.}(2022)]%
        {ran2022surface}
\bibfield{author}{\bibinfo{person}{Haoxi Ran} {et~al\mbox{.}}}
  \bibinfo{year}{2022}\natexlab{}.
\newblock \showarticletitle{Surface representation for point clouds}. In
  \bibinfo{booktitle}{\emph{Proceedings of the IEEE/CVF Conference on Computer
  Vision and Pattern Recognition (CVPR)}}. \bibinfo{pages}{18942--18952}.
\newblock


\bibitem[Ronneberger et~al\mbox{.}(2015)]%
        {ronneberger2015u}
\bibfield{author}{\bibinfo{person}{Olaf Ronneberger} {et~al\mbox{.}}}
  \bibinfo{year}{2015}\natexlab{}.
\newblock \showarticletitle{U-net: Convolutional networks for biomedical image
  segmentation}. In \bibinfo{booktitle}{\emph{International Conference on
  Medical image computing and computer-assisted intervention}}. Springer,
  \bibinfo{pages}{234--241}.
\newblock


\bibitem[Rukhovich et~al\mbox{.}(2022)]%
        {rukhovich2022fcaf3d}
\bibfield{author}{\bibinfo{person}{Danila Rukhovich} {et~al\mbox{.}}}
  \bibinfo{year}{2022}\natexlab{}.
\newblock \showarticletitle{FCAF3D: fully convolutional anchor-free 3D object
  detection}. In \bibinfo{booktitle}{\emph{Proceedings of the European
  conference on computer vision (ECCV)}}. Springer, \bibinfo{pages}{477--493}.
\newblock


\bibitem[Sandler et~al\mbox{.}(2018)]%
        {Sandler_2018_CVPR}
\bibfield{author}{\bibinfo{person}{Mark Sandler} {et~al\mbox{.}}}
  \bibinfo{year}{2018}\natexlab{}.
\newblock \showarticletitle{MobileNetV2: Inverted Residuals and Linear
  Bottlenecks}. In \bibinfo{booktitle}{\emph{Proceedings of the IEEE/CVF
  Conference on Computer Vision and Pattern Recognition (CVPR)}}.
\newblock


\bibitem[Seo et~al\mbox{.}(2021)]%
        {10.1145/3460352}
\bibfield{author}{\bibinfo{person}{Wonik Seo} {et~al\mbox{.}}}
  \bibinfo{year}{2021}\natexlab{}.
\newblock \showarticletitle{SLO-Aware Inference Scheduler for Heterogeneous
  Processors in Edge Platforms}.
\newblock \bibinfo{journal}{\emph{ACM Trans. Archit. Code Optim.}}
  \bibinfo{volume}{18}, \bibinfo{number}{4}, Article \bibinfo{articleno}{43}
  (\bibinfo{date}{jul} \bibinfo{year}{2021}), \bibinfo{numpages}{26}~pages.
\newblock
\showISSN{1544-3566}


\bibitem[Shen et~al\mbox{.}(2020)]%
        {shen2020q}
\bibfield{author}{\bibinfo{person}{Sheng Shen} {et~al\mbox{.}}}
  \bibinfo{year}{2020}\natexlab{}.
\newblock \showarticletitle{Q-bert: Hessian based ultra low precision
  quantization of bert}. In \bibinfo{booktitle}{\emph{Proceedings of the AAAI
  Conference on Artificial Intelligence}}, Vol.~\bibinfo{volume}{34}.
  \bibinfo{pages}{8815--8821}.
\newblock


\bibitem[Shi et~al\mbox{.}(2020)]%
        {shi2020pv}
\bibfield{author}{\bibinfo{person}{Shaoshuai Shi} {et~al\mbox{.}}}
  \bibinfo{year}{2020}\natexlab{}.
\newblock \showarticletitle{Pv-rcnn: Point-voxel feature set abstraction for 3d
  object detection}. In \bibinfo{booktitle}{\emph{Proceedings of the IEEE/CVF
  Conference on Computer Vision and Pattern Recognition (CVPR)}}.
  \bibinfo{pages}{10529--10538}.
\newblock


\bibitem[Shi et~al\mbox{.}(2021)]%
        {shi2021pv}
\bibfield{author}{\bibinfo{person}{Shaoshuai Shi} {et~al\mbox{.}}}
  \bibinfo{year}{2021}\natexlab{}.
\newblock \showarticletitle{PV-RCNN++: Point-voxel feature set abstraction with
  local vector representation for 3D object detection}.
\newblock \bibinfo{journal}{\emph{arXiv preprint arXiv:2102.00463}}
  (\bibinfo{year}{2021}).
\newblock


\bibitem[Song et~al\mbox{.}(2015)]%
        {song2015sun}
\bibfield{author}{\bibinfo{person}{Shuran Song} {et~al\mbox{.}}}
  \bibinfo{year}{2015}\natexlab{}.
\newblock \showarticletitle{Sun rgb-d: A rgb-d scene understanding benchmark
  suite}. In \bibinfo{booktitle}{\emph{Proceedings of the IEEE/CVF Conference
  on Computer Vision and Pattern Recognition (CVPR)}}.
  \bibinfo{pages}{567--576}.
\newblock


\bibitem[Sun et~al\mbox{.}(2020)]%
        {sun-etal-2020-mobilebert}
\bibfield{author}{\bibinfo{person}{Zhiqing Sun} {et~al\mbox{.}}}
  \bibinfo{year}{2020}\natexlab{}.
\newblock \showarticletitle{{M}obile{BERT}: a Compact Task-Agnostic {BERT} for
  Resource-Limited Devices}. In \bibinfo{booktitle}{\emph{Proceedings of the
  58th Annual Meeting of the Association for Computational Linguistics}}.
  \bibinfo{pages}{2158--2170}.
\newblock


\bibitem[Tan et~al\mbox{.}(2019a)]%
        {DBLP:conf/icml/TanL19}
\bibfield{author}{\bibinfo{person}{Mingxing Tan} {et~al\mbox{.}}}
  \bibinfo{year}{2019}\natexlab{a}.
\newblock \showarticletitle{EfficientNet: Rethinking Model Scaling for
  Convolutional Neural Networks}. In \bibinfo{booktitle}{\emph{Proceedings of
  the 36th International Conference on Machine Learning}}.
\newblock


\bibitem[Tan et~al\mbox{.}(2019b)]%
        {Tan_2019_CVPR}
\bibfield{author}{\bibinfo{person}{Mingxing Tan} {et~al\mbox{.}}}
  \bibinfo{year}{2019}\natexlab{b}.
\newblock \showarticletitle{MnasNet: Platform-Aware Neural Architecture Search
  for Mobile}. In \bibinfo{booktitle}{\emph{Proceedings of the IEEE/CVF
  Conference on Computer Vision and Pattern Recognition (CVPR)}}.
\newblock


\bibitem[Tan et~al\mbox{.}(2020)]%
        {Tan_2020_CVPR}
\bibfield{author}{\bibinfo{person}{Mingxing Tan} {et~al\mbox{.}}}
  \bibinfo{year}{2020}\natexlab{}.
\newblock \showarticletitle{EfficientDet: Scalable and Efficient Object
  Detection}. In \bibinfo{booktitle}{\emph{Proceedings of the IEEE/CVF
  Conference on Computer Vision and Pattern Recognition (CVPR)}}.
\newblock


\bibitem[Vora et~al\mbox{.}(2020)]%
        {vora2020pointpainting}
\bibfield{author}{\bibinfo{person}{Sourabh Vora}, \bibinfo{person}{Alex~H
  Lang}, \bibinfo{person}{Bassam Helou}, {and} \bibinfo{person}{Oscar
  Beijbom}.} \bibinfo{year}{2020}\natexlab{}.
\newblock \showarticletitle{Pointpainting: Sequential fusion for 3d object
  detection}. In \bibinfo{booktitle}{\emph{Proceedings of the IEEE/CVF
  Conference on Computer Vision and Pattern Recognition (CVPR)}}.
  \bibinfo{pages}{4604--4612}.
\newblock


\bibitem[Vu et~al\mbox{.}(2022)]%
        {vu2022softgroup}
\bibfield{author}{\bibinfo{person}{Thang Vu} {et~al\mbox{.}}}
  \bibinfo{year}{2022}\natexlab{}.
\newblock \showarticletitle{Softgroup for 3d instance segmentation on point
  clouds}. In \bibinfo{booktitle}{\emph{Proceedings of the IEEE/CVF Conference
  on Computer Vision and Pattern Recognition (CVPR)}}.
  \bibinfo{pages}{2708--2717}.
\newblock


\bibitem[Wang et~al\mbox{.}(2022a)]%
        {wang2022cagroup3d}
\bibfield{author}{\bibinfo{person}{Haiyang Wang} {et~al\mbox{.}}}
  \bibinfo{year}{2022}\natexlab{a}.
\newblock \showarticletitle{CAGroup3D: Class-Aware Grouping for 3D Object
  Detection on Point Clouds}.
\newblock \bibinfo{journal}{\emph{arXiv preprint arXiv:2210.04264}}
  (\bibinfo{year}{2022}).
\newblock


\bibitem[Wang et~al\mbox{.}(2022b)]%
        {wang2022rbgnet}
\bibfield{author}{\bibinfo{person}{Haiyang Wang} {et~al\mbox{.}}}
  \bibinfo{year}{2022}\natexlab{b}.
\newblock \showarticletitle{Rbgnet: Ray-based grouping for 3d object
  detection}. In \bibinfo{booktitle}{\emph{Proceedings of the IEEE/CVF
  Conference on Computer Vision and Pattern Recognition (CVPR)}}.
  \bibinfo{pages}{1110--1119}.
\newblock


\bibitem[Wang et~al\mbox{.}(2017)]%
        {10.1145/3072959.3073608}
\bibfield{author}{\bibinfo{person}{Peng-Shuai Wang} {et~al\mbox{.}}}
  \bibinfo{year}{2017}\natexlab{}.
\newblock \showarticletitle{O-CNN: Octree-Based Convolutional Neural Networks
  for 3D Shape Analysis}.
\newblock \bibinfo{journal}{\emph{ACM Trans. Graph.}} \bibinfo{volume}{36},
  \bibinfo{number}{4}, Article \bibinfo{articleno}{72} (\bibinfo{date}{July}
  \bibinfo{year}{2017}), \bibinfo{numpages}{11}~pages.
\newblock
\showISSN{0730-0301}


\bibitem[Wang et~al\mbox{.}(2020)]%
        {8963950}
\bibfield{author}{\bibinfo{person}{Siqi Wang} {et~al\mbox{.}}}
  \bibinfo{year}{2020}\natexlab{}.
\newblock \showarticletitle{Neural Network Inference on Mobile SoCs}.
\newblock \bibinfo{journal}{\emph{IEEE Design Test}} \bibinfo{volume}{37},
  \bibinfo{number}{5} (\bibinfo{year}{2020}), \bibinfo{pages}{50--57}.
\newblock


\bibitem[Wang et~al\mbox{.}(2022c)]%
        {wang2022multimodal}
\bibfield{author}{\bibinfo{person}{Yikai Wang} {et~al\mbox{.}}}
  \bibinfo{year}{2022}\natexlab{c}.
\newblock \showarticletitle{Multimodal token fusion for vision transformers}.
  In \bibinfo{booktitle}{\emph{Proceedings of the IEEE/CVF Conference on
  Computer Vision and Pattern Recognition (CVPR)}}.
  \bibinfo{pages}{12186--12195}.
\newblock


\bibitem[Xu et~al\mbox{.}(2021)]%
        {9525229}
\bibfield{author}{\bibinfo{person}{Zhiyuan Xu} {et~al\mbox{.}}}
  \bibinfo{year}{2021}\natexlab{}.
\newblock \showarticletitle{A Co-Scheduling Framework for DNN Models on Mobile
  and Edge Devices with Heterogeneous Hardware}.
\newblock \bibinfo{journal}{\emph{IEEE Transactions on Mobile Computing}}
  (\bibinfo{year}{2021}), \bibinfo{pages}{1--1}.
\newblock
\urldef\tempurl%
\url{https://doi.org/10.1109/TMC.2021.3107424}
\showDOI{\tempurl}


\bibitem[Yang et~al\mbox{.}(2022)]%
        {yang2022boosting}
\bibfield{author}{\bibinfo{person}{Hao Yang} {et~al\mbox{.}}}
  \bibinfo{year}{2022}\natexlab{}.
\newblock \showarticletitle{Boosting 3D Object Detection via Object-Focused
  Image Fusion}.
\newblock \bibinfo{journal}{\emph{arXiv preprint arXiv:2207.10589}}
  (\bibinfo{year}{2022}).
\newblock


\bibitem[Ye et~al\mbox{.}(2020)]%
        {Ye_2020_CVPR}
\bibfield{author}{\bibinfo{person}{Maosheng Ye} {et~al\mbox{.}}}
  \bibinfo{year}{2020}\natexlab{}.
\newblock \showarticletitle{HVNet: Hybrid Voxel Network for LiDAR Based 3D
  Object Detection}. In \bibinfo{booktitle}{\emph{Proceedings of the IEEE/CVF
  Conference on Computer Vision and Pattern Recognition (CVPR)}}.
\newblock


\bibitem[Yi et~al\mbox{.}(2020)]%
        {yi2020heimdall}
\bibfield{author}{\bibinfo{person}{Juheon Yi}, \bibinfo{person}{},
  {et~al\mbox{.}}} \bibinfo{year}{2020}\natexlab{}.
\newblock \showarticletitle{Heimdall: mobile GPU coordination platform for
  augmented reality applications}. In \bibinfo{booktitle}{\emph{Proceedings of
  the 26th Annual International Conference on Mobile Computing and
  Networking}}. \bibinfo{pages}{1--14}.
\newblock


\bibitem[Zhang et~al\mbox{.}(2020)]%
        {zhang2020h3dnet}
\bibfield{author}{\bibinfo{person}{Zaiwei Zhang} {et~al\mbox{.}}}
  \bibinfo{year}{2020}\natexlab{}.
\newblock \showarticletitle{H3dnet: 3d object detection using hybrid geometric
  primitives}. In \bibinfo{booktitle}{\emph{Proceedings of the European
  conference on computer vision (ECCV)}}. Springer, \bibinfo{pages}{311--329}.
\newblock


\bibitem[Zhou et~al\mbox{.}(2018)]%
        {Zhou_2018_CVPR}
\bibfield{author}{\bibinfo{person}{Yin Zhou} {et~al\mbox{.}}}
  \bibinfo{year}{2018}\natexlab{}.
\newblock \showarticletitle{VoxelNet: End-to-End Learning for Point Cloud Based
  3D Object Detection}. In \bibinfo{booktitle}{\emph{Proceedings of the
  IEEE/CVF Conference on Computer Vision and Pattern Recognition (CVPR)}}.
\newblock


\bibitem[Zhu et~al\mbox{.}(2020)]%
        {zhu2020deformable}
\bibfield{author}{\bibinfo{person}{Xizhou Zhu}, \bibinfo{person}{Weijie Su},
  \bibinfo{person}{Lewei Lu}, \bibinfo{person}{Bin Li},
  \bibinfo{person}{Xiaogang Wang}, {and} \bibinfo{person}{Jifeng Dai}.}
  \bibinfo{year}{2020}\natexlab{}.
\newblock \showarticletitle{Deformable detr: Deformable transformers for
  end-to-end object detection}.
\newblock \bibinfo{journal}{\emph{arXiv preprint arXiv:2010.04159}}
  (\bibinfo{year}{2020}).
\newblock


\end{thebibliography}
